\documentclass[nohyper,letterpaper,11pt,notoc]{JHEP3}
\usepackage{graphicx}
\usepackage{amssymb}
\usepackage{amsmath}
\usepackage{latexsym}
\usepackage[all]{xy}
\usepackage{feynmf}
\usepackage{slashed}
\preprint{}

\newcommand{\ket}[1]{|#1\rangle}
\newcommand{\be}{\begin{equation}}
\newcommand{\ee}{\end{equation}}
\newcommand{\bea}{\begin{eqnarray}}
\newcommand{\eea}{\end{eqnarray}}
\newcommand{\GeV}{~\mathrm{GeV}}
\newcommand{\TeV}{~\mathrm{TeV}}

\title{Spin Measurements in Cascade Decays at the LHC }
\author{Lian-Tao Wang and Itay Yavin \\ Jefferson Physical Laboratory,
  Harvard University, Cambridge, MA 02138}

\abstract{We systematically study the possibility of determining the spin of new particles after their discovery at the
LHC. We concentrate on angular correlations in cascade decays. Motivated by constraints of electroweak precision tests
and the potential of providing a Cold Dark Matter candidate, we focus on scenarios of new physics in which some
discrete symmetry guarantees the existence of stable neutral particles which escape the detector. More specifically, we
compare supersymmetry with another generic scenario in which new physics particles have the same spin as their Standard
Model partners. A survey of possibilities of observing spin correlations in a broad range of decay channels is carried
out, with interesting ones identified. Rather than confining ourselves to one ``collider friendly" benchmark point
(such as SPS1a), we describe the parameter region in which any particular decay channel is effective. We conduct a more
detailed study of chargino's spin determination in the decay channel $\tilde{q}\rightarrow q + \tilde{C}^\pm
\rightarrow q + W^\pm + LSP$. A scan over the chargino and neutralino masses is performed. We find that as long as the
spectrum is not too degenerate the prospects for spin determination in this channel are rather good.}

\begin{document}
\section{Introduction}

Naturalness of the weak scale implies the existence of new physics beyond the Standard Model at the scale $\sim \TeV$.
Typical new physics scenarios predict the existence of a set of new particles at that scale. The Large Hadron Collider
(LHC) gives us a great opportunity for discovering those particles.

In order to understand the nature of the new physics, it is necessary to measure its properties in detail. One of the
obvious tasks is to reconstruct the masses \cite{Baer:1995nq,Baer:1995va,Hinchliffe:1996iu,Hinchliffe:1998zj,unknown:1999fr,Lester:1999tx,Lester:2005je,Miller:2005zp} and gauge quantum numbers of the new particles from experimental
data. A recent study \cite{Arkani-Hamed:2005px} demonstrated the challenges of such a goal and suggested possible
directions in achieving it.

On the other hand, there is another, at least equally important, LHC inverse problem: how do we determine the spin of
any newly discovered particle? The proposed new particles in several main candidates of new physics scenarios typically
have similar gauge interactions. They could often be organized as partners of the known Standard Model particles with
the same gauge quantum number, such as quark partners, lepton partners and gauge boson partners. A typical example is
the set of superpartners in supersymmetry, including squarks, sleptons, gauginos, and so on. Another interesting
scenario is the theory space models inspired by \cite{Arkani-Hamed:2001ca}.  The duplication of the Standard Model
states in this scenario comes from introducing more copies of the Standard Model gauge group.  Typical examples are
little Higgs models \footnote{Another well-known example is the the extra-dimensional setup with the corresponding KK
particles. As we learned in the past few years, this is related to the theory space models via deconstruction}.
Measuring the partners' spin becomes a crucial, sometimes single, way to distinguish those scenarios.

Motivated by electroweak precision constraints and the existence of Cold Dark Matter, many new physics scenarios
incorporate some discrete symmetry which guarantees the existence of a lightest stable neutral particle, LSNP.
Well-known examples of such discrete symmetries include R-parity in supersymmetry, KK-parity in universal
extra-dimension models \cite{Appelquist:2000nn}, or similarly, T-parity in Little Higgs Models
\cite{Cheng:2003ju,Cheng:2004yc,Low:2004xc,Cheng:2005as}. The existence of such a neutral particle at the end of the
decay chain results in large missing energy events in which new physics particles are produced. This fact helps to
separate them from the Standard Model background. On the other hand, it also makes the spin measurement more
complicated because it is almost impossible to reconstruct the momentum, and therefore the rest frame, of the decaying
particles.

The question of spin determination has been revisited recently. The total cross section might serve as an initial hint
to the spin of the new particles discovered \cite{Datta:2005vx}. This is not entirely satisfactory because certain
model dependence is inevitable when using the rate information. For example, a fermion can be faked by two closely degenerate scalars.
Moreover, such a determination is only possible if we could measure the masses of the particle using kinematical
information. As demonstrated in \cite{Cheng:2005as,Meade:2006dw}, typical ``transverse'' kinematical observables are
not sensitive to the absolute mass of particles. One can only deduce the mass difference between the decaying particle
and the neutral particle escaping the detector. With some assumptions regarding the underlaying model there are more
subtle kinematical observables which, in combination with the rate information, could determine the spin
\cite{Meade:2006dw}. To what extent this could be generalized to a broader classes of new physics particles is
currently under investigation.

Therefore, it is important to investigate other possible ways of directly measuring the spin of new particles. The
typical way of measuring the spin of a decaying particle starts with reconstructing its rest frame from the decay
products. Then, the angular distribution in the rest frame contains the full spin information, independent of the
boost. As discussed above, we do not have enough kinematical information to boost to the rest frame of the decaying
particle if the spectrum contains a LSNP. Therefore, it is natural to consider distributions as a function of
relativistic invariants constructed out of the decay products of a single decaying particle. We will focus on this
possibility in this paper.

Various new physics models always have some detailed differences in their spectra. But, such differences are very model
dependent. Although in principle they could carry interesting information, we will focus on spin determination based on
Standard Model partners only. What we have in mind are two classes of models with almost identical gauge quantum
numbers and maximal flexibility in their mass spectra. In other words, they would look very similar except for their
spin content.

One obvious scenario is low energy supersymmetry, parameterized by the MSSM with a conserved R-parity.

As a contrasting scenario we consider a framework in which all new particles have the same spin as their Standard Model
counter parts. The existence of a LSNP, is guaranteed by the assumption that all the new physics particles are odd
under a certain $Z_2$ parity. Special cases of this scenario could be the first KK level of UED or T-parity little
Higgs. However, what we have in mind is a more generic setup and we will not constrain ourselves to any special mass or
coupling relations imposed by these two scenarios. To emphasize its generic nature, we will call it the Same Spin
scenario in this paper. We will use symbols with primes to label the new particles in the Same Spin scenario. For
example, we will use $q'$ to label the quark partner, and so on. We will assume the LSNP in this case is a vector and
label it as $A'$.

Typical new physics scenarios have many complicated decay channels. Many kinematical distributions can be constructed
from them. One of the main goals of this paper is to present a systematic survey of the observability of spin
correlations in a wide variety of decay chains which are generically present in new physics scenarios. We identify
interesting decay channels to focus on for spin measurements. It is important to notice, as will be clear from our
discussion, that the usefulness of any particular channel is restricted to a specific range of parameters. We describe
the kinematical requirements for each of the channels we analyze. There is no obvious golden channel. For different
points in the parameter space, we will generically have different decay channels available. Therefore, we will have to
devise different strategies depending on mass spectra of underlying models.

One of the important tools we develop in this paper is a set of simple rules, which summarizes many well-known results
concerning spin correlations. Such simple rules allows one to gain insight into the angular correlations in decay,
without the necessity of going through a lengthy calculation. They can be useful in other, potentially more
complicated, scenarios than the ones considered hereafter.

Barr \cite{Barr:2004ze} investigated a typical supersymmetry cascade involving a squark decay. He found that angular
correlations exist between the decay products. Barr's method relies on the fact that squarks and anti-squarks are
produced unevenly in a proton-proton collider. There are several follow-up studies along the same lines
\cite{Datta:2005zs,Smillie:2005ar,Alves:2006df}. References \cite{Smillie:2005ar} and \cite{Datta:2005zs} went further
and contrasted supersymmetry with the universal extra-dimensions scenario \cite{Appelquist:2000nn},\cite{Cheng:2002ab}.
Reference \cite{Smillie:2005ar} found that with a mass spectrum given by the SPS point 1a, the SUSY model is
distinguishable from the UED case. In their study, they assumed the lepton can be perfectly correlated with the correct
jet. That might be possible if complete kinematic information is available, but in practice seems quite difficult. In
general jet combinatorics must be taken into account.

One limitation of these investigations is the need for a light leptonic partner. It must be lighter than the second
lightest neutral gauge boson partner, such as the second lightest neutralino (which must be a wino or bino), in
supersymmetry. While true in some special benchmark models \cite{Allanach:2002nj}, there is no reason to assume this is
a generic feature of supersymmetry breaking. In fact, it is more generic to assume otherwise, especially if one is
driven by the problem of naturalness.

We present a detailed study of spin correlations in the decay chain $\tilde{q} \rightarrow q + \tilde{C}_1^{\pm}
\rightarrow q+W^{\pm} + \tilde{N}_1 $ in supersymmetry and its counter part $q' \rightarrow q + W^{\prime \pm}
\rightarrow q + W^{\pm} + A'$ in the Same Spin scenario.  Such a decay chain does not require the leptonic partner to
be lighter than the gauge boson partner and is certainly more generic in parameter space. We will assume that the mass
splitting between the two lightest states is greater than $m_{\rm W,Z}$. Therefore, the on-shell decay to W/Z always
dominates\footnote{If the mass splitting is less than $m_{\rm W,Z}$, sometimes, the off-shell diagram via a squark and
slepton can be important. Although it is a special case of the mass spectrum, it is certainly worthwhile exploring it
further.}. Our result shows that it is possible to observe spin correlation in this decay chain. As a demonstration of
the result of our general discussion, we map out the parameter region in which this decay channel is useful.

As part of the analysis we used HERWIG 6.507 \cite{Corcella:2002jc} which implements a spin correlation algorithm. This
algorithm was first used for QCD parton showers \cite{Collins:1987cp,Knowles:1987cu,Knowles:1988hu,Knowles:1988vs} and
later extended by Richardson to supersymmetric and top processes \cite{Richardson:2001df}. We supplemented the code to
include the decays of massive gauge bosons (such as KK partners of the gluon) and the details are spelled out in
appendix \ref{app: HERWIG}. To the best of the authors' knowledge, HERWIG is the only simulator that implements such an
algorithm and is therefore suitable for spin determination studies.

The paper is organized as follows: In Section~\ref{sec: intuition} we try to build some intuition by looking at the
effects of spin on the angular distributions of simple decays. In Section~\ref{sec:survey}, we present a survey of
spin correlations in various decay channels. Our detailed study of the decay chain with $q W^{\pm}$ final states is
presented in Section~\ref{sec: noleptons}. There, we take up the task of constructing an observable signal to
distinguish SUSY from  the Same Spin scenario in the absence of any leptonic partners. Finally, in section~\ref{sec:
Future}, we comment on possible future directions and present our conclusions.

\section{Simple Spin Correlations}
\label{sec: intuition}

In this section we review some basic angular distributions from simple decays. These distributions will serve as
building blocks in our understanding of the spin correlations in more complicated decay chains which we will consider
later.

\subsection{Scalar decay}
A scalar does not pick any special direction in space and so its decay is isotropic. It does not mean that the
existence of scalars spoils any hope for distinguishing them away from phase-space. The production of bosons (via a
$Z^0$ for example) has a different angular distribution about the beam axis than that of fermions. This discrepancy can
be employed in determining the spin of lepton partners (see for example, \cite{Barr:2005dz}). However, in our study we
will concentrate on a single branch in which case it is not possible to distinguish a scalar from phase-space.

\subsection{Fermion decay}

First, we consider the decay of a fermion $\psi_1$ into another fermion $\psi_2$ and a scalar $\phi$, via an
interaction of the form
\begin{equation}
\label{f1f2phi} y_L \phi \bar{\psi}_2 P_L \psi_1 +y_R \phi \bar{\psi}_2 P_R \psi_1
\end{equation}
Depending on the model, this coupling could be either chiral, $y_L \neq y_R$, or non-chiral,  $y_L = y_R$. We will see
examples of both cases in our study.

\begin{figure}[h]
\begin{center}
\includegraphics[scale=0.8]{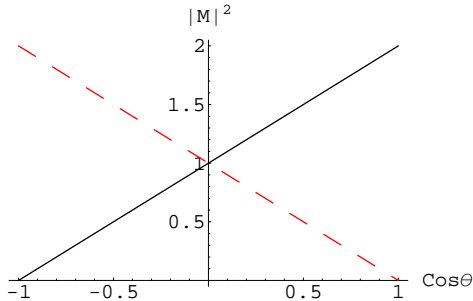}
\end{center}
\caption{The decay probability for a fermion into a scalar and another fermion of the same helicity (solid-black) or
opposite helicity (dashed-red) as a function of  $\cos\theta$. $\theta$ is defined with respect to the axis of
polarization of the decaying fermion.} \label{fig: fermionDecay}
\end{figure}

If the coupling in Eq.~\ref{f1f2phi} is chiral, $\psi_2$ is produced in a chirality eigenstate.  If $\psi_2$ is boosted
then it is in a helicity eigenstate, i.e., polarized. However, $\psi_1$ is, in general, not polarized and therefore the
decay is isotropic, even if the coupling (\ref{f1f2phi}) is chiral and $\psi_2$ is boosted. It is easy to see how this
comes about. If it is a Left handed coupling, $y_R=0$, then $\psi_2$ is mostly a right-handed particle,
$\ket{\downarrow}$. From the transformation of a spinor under a rotation by an angle $\theta$ we have that,
\begin{eqnarray*}
\ket{\uparrow} &\rightarrow&  \cos\left(\displaystyle\frac{\theta}{2}\right) ~\ket{\uparrow} +
\sin\left(\frac{\theta}{2}\right)
~\ket{\downarrow} \\
\ket{\downarrow} &\rightarrow&  -\sin\left(\displaystyle\frac{\theta}{2}\right) ~\ket{\uparrow} +
\cos\left(\frac{\theta}{2}\right) ~\ket{\downarrow}
\end{eqnarray*}
The angle $\theta$ is defined with respect to $\psi_1$ polarization axis. Notice that if $\psi_1$ is left-handed
polarized, $\ket{\uparrow}$, its decay probability is $\propto \sin^2\left(\frac{\theta}{2}\right)$. On the other hand,
if it is right-handed polarized, $\ket{\downarrow}$, its decay probability $\propto
\cos^2\left(\frac{\theta}{2}\right)$. These decay distributions are shown in Fig.(\ref{fig: fermionDecay}) as a
function of $\cos\left(\theta\right)$. Unfortunately, $\psi_1$ itself is normally not polarized and averaging over the
two process the decay is indeed isotropic.

However, if $\psi_1$ came from the decay of another particle and that vertex was chiral then the situation is
different. In that case $\psi_1$ is polarized and its subsequent decay is governed by a non-trivial angular
distribution as shown in Fig. (\ref{fig: fermionDecay}). Whether the decay involves a helicity flip or not determines
the sign of the slope.

Next, we consider  the decay of a fermion into another fermion and a gauge-boson via an interaction of the form
\begin{equation}
 g_L \bar{\psi}_2 \gamma^{\mu} P_L \psi_1 A_{\mu}+g_R \bar{\psi}_2 \gamma^{\mu} P_R \psi_1 A_{\mu}
\end{equation}
As before, we consider the case where $\psi_2$ is boosted. If the interaction is chiral $\psi_2$ is in a definite
helicity state. The fermionic current that couples to $A^{\mu}$ is of the form $\bar{\psi}_{\dot{\alpha}}
\sigma^{\dot{\alpha}\beta}_\mu \psi_\beta$. If the emitted gauge-boson is longitudinally polarized the distributions
are the same as the decay into a fermion and a scalar. If it transversely polarized it is precisely opposite (i.e. same
helicity corresponds to $\sin^2\theta/2$ and opposite helicity to $\cos^2\theta/2$).

The most important feature of the fermion's decay is the linear dependence of the decay probability on $\cos\theta$. It
is also clear that chiral vertices must be involved in order to observe spin correlations (unless the fermion is a
Majorana particle, a possibility we discuss below).

\subsection{Gauge-boson decay}

\begin{figure}[h]
\begin{center}
\includegraphics[scale=0.7]{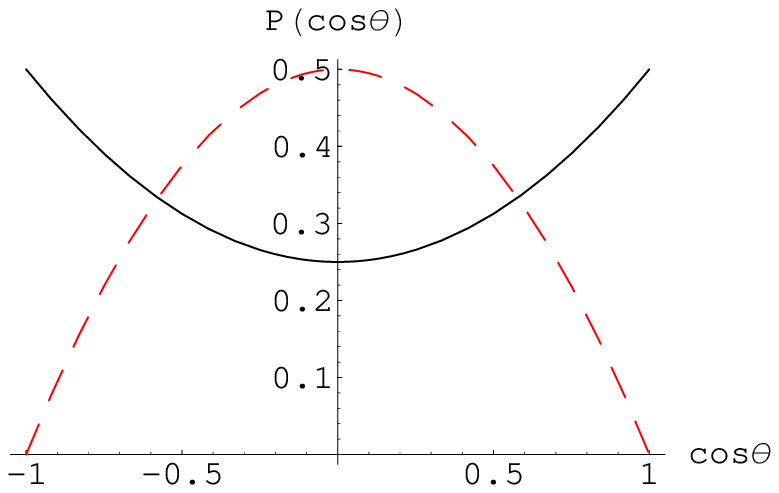} \hspace{.5 in}
\includegraphics[scale=0.7]{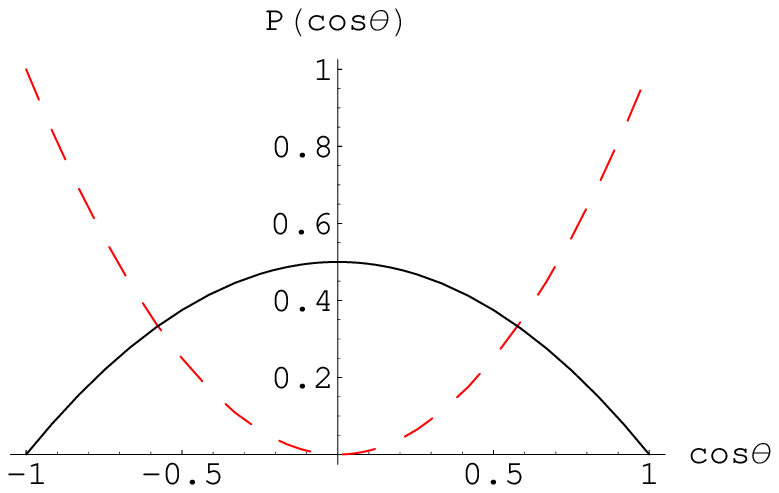}
\end{center}
\caption{The decay probability for a gauge-boson into two fermions (left) and two bosons (right) for transverse
(solid-black) and longitudinal polarization (dashed-red) as a function of  $\cos\theta$.}\label{fig: gbDecayFers}
\end{figure}

When a gauge-boson decay (2-body), relativity forces the products to be two bosons or two fermions. As is well known,
when the products are two fermions the angular distribution is given by,
\begin{equation}
\label{eqn:MgbPolFer} P_{\rm trans} (\cos\theta) = \frac{1}{4}\left(1 + \cos^2\theta\right) \quad \quad    P_{\rm long}
(\cos\theta) =  \frac{1}{2}\left(1 - \cos^2\theta\right)
\end{equation}
If a gauge boson decays into two scalars via the interaction
\begin{equation} g \phi_2^*
\stackrel{\leftrightarrow}{\partial}_{\mu} \phi_1 A^{\mu} ,
\end{equation}
the angular distribution has the opposite structure,
\begin{equation}
\label{eqn:MgbPolBos} P_{\rm trans} (\cos\theta) = \frac{1}{2}(1-\cos^2\theta) \quad \quad    P_{\rm long} (\cos\theta)
= \cos^2\theta
\end{equation}
where the subscript on $P$ denotes the initial gauge-boson's polarization. As usual $\theta$ is defined about the
polarization axis. The decay of a gauge-boson into two other gauge-bosons has the same angular distribution as Eq.
(\ref{eqn:MgbPolBos}). These are shown in Fig.(\ref{fig: gbDecayFers}).
As usual there are finite mass effects that come into play when the products are not highly boosted. Those tend to wash
out any angular dependence of the amplitude. Generically these contributions scale as $m^2/E^2$. Therefore, as noted
before there has to be an appreciable difference between the mass of the decaying particle and its products so that
$m^2/E^2 \lesssim 1/2$.

The contrast with the previous case is clear as the dependence of the amplitude on $\cos\theta$ is quadratic. It is
also important to note that the vertex need not be chiral.

\subsection{Higher spin}

By noting that a rotation by $\theta$ of a state of spin $j$ is given
by $e^{i\theta j\sigma_y}$ it is easy to see that the amplitude for
the decay of a particle with spin $j$ is some polynomial of degree
$2j$,
\begin{equation}
  P_\lambda(\cos\theta) = a_{2j}(\cos\theta)^{2j} +
  a_{2j-1}(\cos\theta)^{2j-1} + \ldots + a_0
\end{equation}
The coefficients $a_i$ are such that when we sum over all
polarizations $\lambda$ we get,
\begin{equation}
  \sum_\lambda P_\lambda(\cos\theta) = 1
\end{equation}
since an unpolarized particle has no preferred direction. In this paper we concentrate on spin 0,1/2, and 1 and will
not consider higher spin. Nonetheless, this is an important issue to address. For example, if the partners of the
graviton are indeed detected it would be good to know whether it is a supersymmetric spin-3/2 object or a Same-Spin
spin-2 resonance.

\section{Angular correlations in cascade decays}
\label{sec:survey}

In this section, we present a systematic study of spin correlations in a wide variety of cascade decay channels. Aside
from the matrix element, the kinematics also play a crucial role in the observability of spin effects. We lay out the
conditions for observing spin correlations in each of the decay channels we discuss. Whether any of the channels is
open or not depends on the particular mass spectrum. However, it is not unreasonable to expect several such channels to
be open in a generic model. This is important because the signal from any one channel might not be sufficiently strong.
In this case we would have to combine the signal from a few channels to obtain a high confidence spin determination.

We focus on a class of specific kinematical observables. It is constructed from the momenta of two of
the observed final state particles. More complicated decay patterns and observables consisting of more than two
observable particles could also be interesting.

A generic feature of this type of observables is that spin information of the {\it intermediate} particle, which has
observable decay products on both sides, in the decay chain always manifest itself as some polynomial structure in the
distribution. Indeed, a particle of spin $j$, if polarized, will result in a polynomial of degree $2j$. On the other
hand, such a method is not useful for determining the spin of any particle at the top or bottom of the decay chain. For
the same reason, very short decay chains such as $\tilde{q} \rightarrow q + {\rm LSP}$ won't contain much information.

As discussed in Section~\ref{sec: intuition}, a key requirement for the existence of any spin correlations is for the
intermediate state particle to be polarized\footnote{Strictly speaking, this requirement only applies for on-shell
particles. For off-shell particles, the spin correlation could have new interesting properties. We will examine the
off-shell decay in Section~\ref{subsec: Offshell} }. A boost invariant way to know whether a particle is polarized or
not is to study this question in its rest frame using a direction defined by its mother particle and the other decay
products. We will see examples of such analysis in the decay channels we consider below. From the discussion in the
previous section, it is clear that there are only a few ways for a particle to be polarized in its rest frame,
\begin{enumerate}
\item For Majorana fermions a spin flip results in a different process with different end products. We must be able to
tell those apart (measuring a leptons vs. anti-lepton). We will see this in detail in the discussion to follow.
\item Dirac fermions must be produced from a (partially) chiral coupling and decay through a (partially) chiral interaction.
\item In general the spectrum of new particles needs not be left-right symmetric. In this case, the interaction
of these particles is effectively chiral even if the gauge-coupling is vector like. A typical example is the asymmetric
QCD production of left and right squarks when their masses are very different\footnote{We would like to thank M. Peskin
for bringing this fact to our attention.}.
\item For a gauge boson, it must come from the decay of a boosted particle (in the gauge-boson rest frame).
\end{enumerate}
We will see detailed realizations of all of these requirements in various decay channels we study in this section.

We will organize our discussion in terms of different final states.

\subsection{Weak Decay with $q \ell^{\pm}$ final state}

We will go through the logic of establishing spin correlation this channel in more detail because many other channels
can be understood following very similar arguments.

We first consider the supersymmetric case. Here, the decay to $q \ell^{\pm}$ final states will proceed through either
Dirac or Majorana fermion intermediate states. We consider the case of a Dirac fermion first (i.e., chargino
intermediate state) and compare it with the corresponding process in the Same Spin scenario where the intermediate
state is a massive gauge boson $W'$. We will comment on the case of Majorana fermion briefly. For more details, see
\cite{Barr:2004ze}.

In the case of a Dirac fermion there are two possible ways for the particle to be polarized. If it is off-shell then
one of the helicities dominates over the other simply because $m^2/q^2\ne 1$, where $q$ is the fermion 4-momenta (this
possibility is also open to a Majorana fermion). This might become important if a decay must proceed through an
off-shell particle simply because no other channel is available. We will not pursue this possibility further, but we
comment on it in section (\ref{subsec: Offshell}).

\begin{figure}[h]
\begin{center}
\includegraphics[scale=0.5]{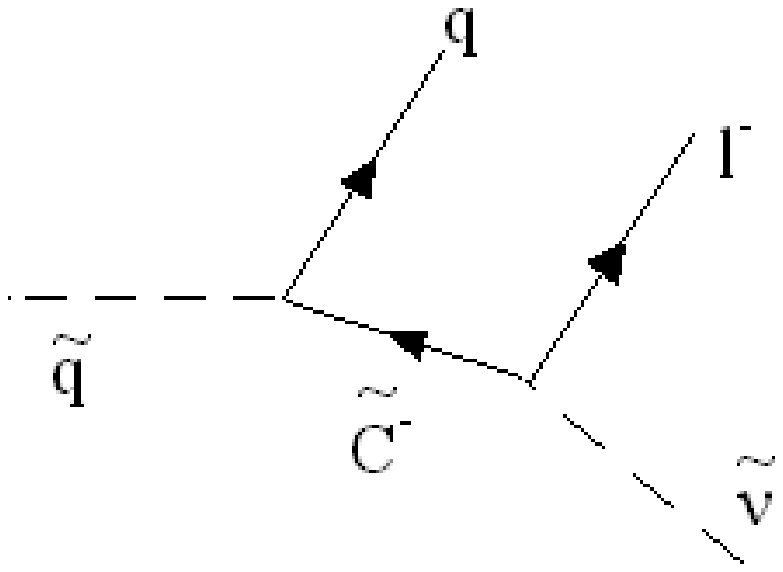} \hspace{.5 in}
\includegraphics[scale=0.8]{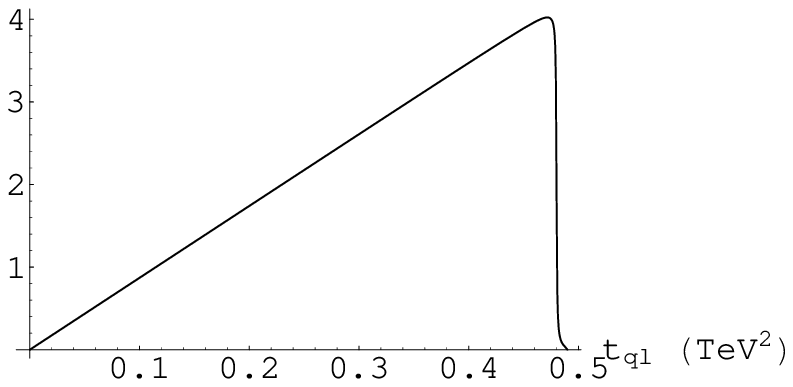}
\end{center}
\caption{The decay of a squark through a chargino involves two chiral vertices. As a result the lepton's direction is
correlated with that of the outgoing quark. On the right we plot the amplitude as a function of $t_{ql}$ for
$m_{\tilde{q}}=1000\GeV$, $m_{\tilde{C}}=500\GeV$ and $m_{\tilde{\nu}}=300\GeV$. The graph is normalized to unit area.}
\label{fig: qkChrgLep}
\end{figure}

The other possibility for a Dirac fermion to be polarized is when both its mother vertex and its daughter vertex are
at least partially chiral. As an example, consider the decay of a squark into a quark, slepton and anti-lepton through
a Chargino, as shown in Fig.(\ref{fig: qkChrgLep}). Using the rules we developed in the previous section it is
straightforward to understand what angular correlations are expected.

In the rest frame of the Chargino, the decaying squark and outgoing quark define a polarization axis. Since the
interaction is chiral the Chargino is polarized. Since the second vertex is also polarized, we have a polarized fermion
decaying into another fermion (lepton) and a scalar (slepton). As we saw before, this decay is governed by a first
order polynomial of $\cos\theta$. However, notice that $\cos\theta$ is related to the relativistically invariant
quantity, $t_{ql}$,
\begin{equation}
t_{ql} = (p_q+p_l)^2 = 2 \frac{\left(m_{\tilde{q}}^2 - m_{\tilde{C}}^2\right) \left(m_{\tilde{C}}^2 -
m_{\tilde{l}}^2\right)}{4m_{\tilde{C}}^2}(1-\cos\theta)
\end{equation}
where the last equality only holds in the Chargino's rest frame.
We can immediately conclude that the relativistically invariant amplitude is at most a linear function of $t_{ql}$ with
the sign given by the explicit details of the couplings. Of course, this can be easily confirmed by an explicit
computation of the amplitude, as shown in Fig.\ref{fig: qkChrgLep}).


In contrast with the supersymmetric case let us consider the decay chain $q' \rightarrow q + W^{\prime \pm} \rightarrow
q + W^{\pm} + \nu'$, where we have assumed that $W'$ couples like a Standard Model $W$. The relevant diagram is shown
in Fig.(\ref{fig: qkKKWLep}). If the spectrum is not too degenerate then in the rest frame of the $W'$, both the
incoming $q'$ and the outgoing $q$ are boosted and mostly left-handed. Therefore, the $W'$ longitudinal polarization
dominates over the transverse one. Another way of seeing the same thing is to note that in the rest frame of $W'$ the
fermionic current can be written in terms of the gauge-boson polarizations,
\begin{equation}
\label{eqn: Wqqkk coupling} g_A\bar{u}_2\gamma^\mu P_Au_1 \propto \left( \epsilon_{\rm long} +
\frac{m_{q'}}{E_{q'}}\left(c_L \epsilon_L + c_R \epsilon_R \right)\right)
\end{equation}
where we have neglected $m_q$. $c_L$ and $c_R$ are $\mathcal{O}(1)$ coefficients depending on the precise nature of the
interaction. Notice the suppression of the transverse polarization with respect to the longitudinal one by a factor of
$m_{q'}/E_{q'}$ in the amplitude. It is clear that when the fermions' mass difference is comparable to the $W^{\prime}$
mass, $m_{q'}-m_q \sim m_{W'}$, the resulting polarization is negligible, since $m_{q'}/E_{q'}\sim 1$ in the rest frame
of the $W^{\prime}$.

 Since the $W^{\prime}$ is longitudinally polarized, its subsequent decay into $l^+ ~ \nu^{\prime}$ is governed by a $1-\cos^2\theta$.
Here, $\theta$ is the angle of the outgoing leptons with respect to the axis of polarization defined by the quarks.
Therefore, the relativistically invariant amplitude squared must be a \textsl{quadratic} function of $t_{ql}$ with a
negative coefficient in front of the leading power.
\begin{figure}[h]
\begin{center}
\includegraphics[scale=0.5]{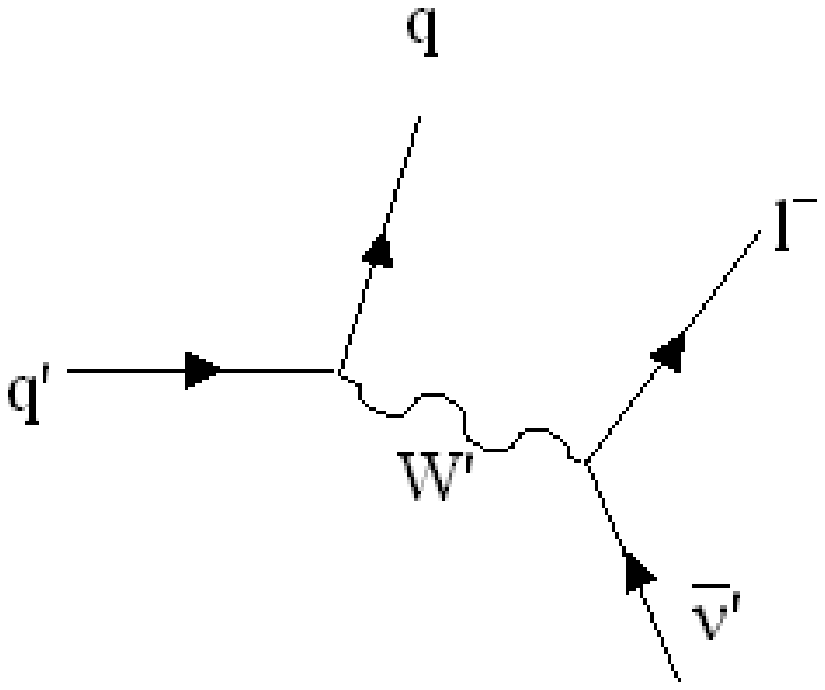} \hspace{.5 in}
\includegraphics[scale=0.9]{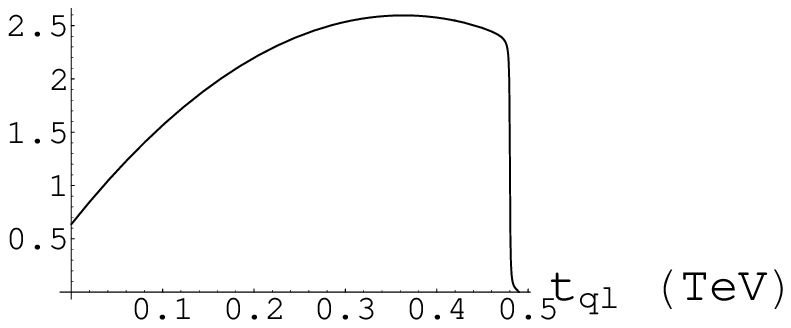}
\end{center}
\caption{When $q'$ decays the intermediate $W^{\prime}$ is longitudinally polarized if the incoming $q^{\prime}$ and
outgoing $q$ are both boosted in its rest frame. This in turn will result in angular correlations between the
directions of the quark and the lepton. On the right we plot the amplitude as a function of $t_{ql}$ for
$m_{q'}=1000\GeV$, $m_{W'}=500\GeV$ and $m_{\nu'}=300\GeV$. The graph is normalized to unit area.} \label{fig:
qkKKWLep}
\end{figure}
Notice that a gauge-boson does not require the vertices to be chiral. This is important and potentially useful in
determining the gluon partner's spin. However, it is also more susceptible to mass difference effects (see equation
(\ref{eqn: Wqqkk coupling})). In contrast, the fermionic counterpart remains polarized even when
$m_{\tilde{q}}$ is not very different from $m_{\tilde{C}}$, as long as the coupling is chiral and the outgoing quark is boosted.

\begin{figure}[h]
\begin{center}
\includegraphics[scale=0.5]{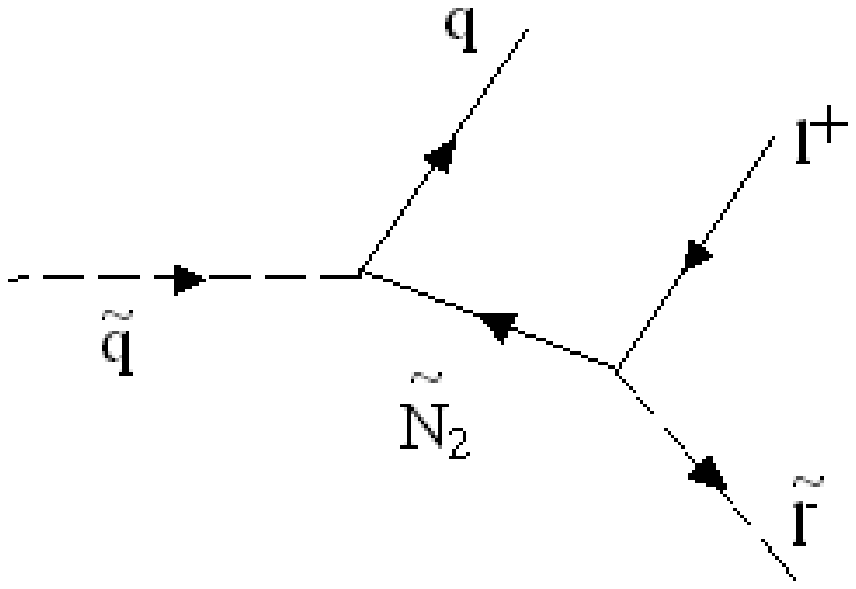} \hspace{.5 in}
\includegraphics[scale=0.5]{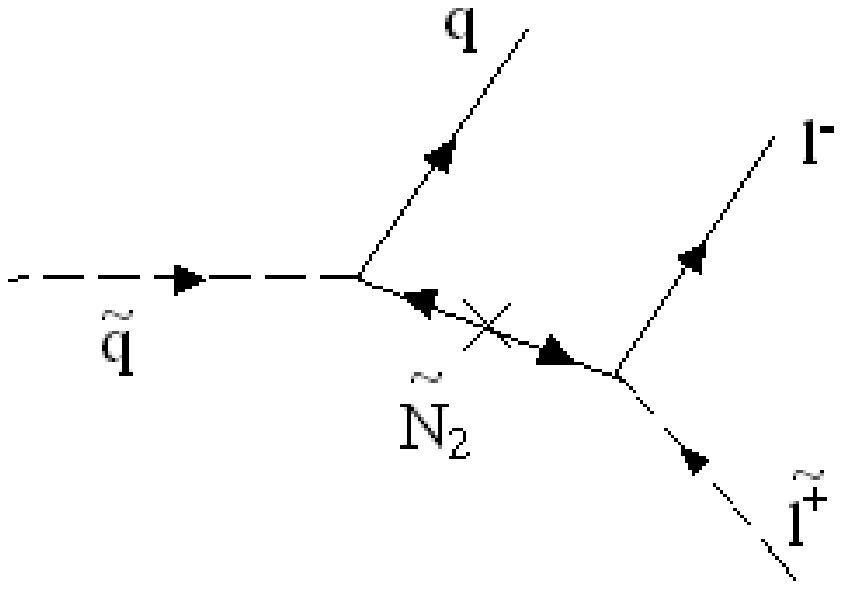}
\end{center}
\caption{The two possible modes for a decay through a Neutralino. A spin flip requires a mass insertion (right), which
results in a different process.} \label{fig: MajDecay}
\end{figure}

Finally, we briefly consider the decay of the squark into $q \ell^{\pm}$ final states via a Majorana fermion
intermediate state. The relevant diagrams are shown in Fig.(\ref{fig: MajDecay}). For the propagator to flip its spin
we must place a mass insertion. However, due to the Majorana nature of the Neutralino, this corresponds to a different
process with different final states than the one without a mass insertion. Therefore, the propagator has a definite
helicity for each of the processes and there are angular correlations between the quark and the lepton. This fact was
exploited by A. Barr~\cite{Barr:2004ze} to determine the spin of the Neutralino and further details can be found in the
reference. One can study $q \ell^+ $ and $q \ell^-$ distributions to uncover the spin information. However, there is a
further complication due the Majorana nature of the Neutralino. There is always another diagram starting from
anti-squark with opposite sign of its charge which contributes to the same process but with the opposite helicity
structure as shown in Fig.(\ref{fig: antiMajDecay}). As Barr noted, in a proton-proton collider squarks and
anti-squarks are produced unevenly and therefore the angular correlations are not washed out completely.

\begin{figure}[h]
\begin{center}
\includegraphics[scale=0.5]{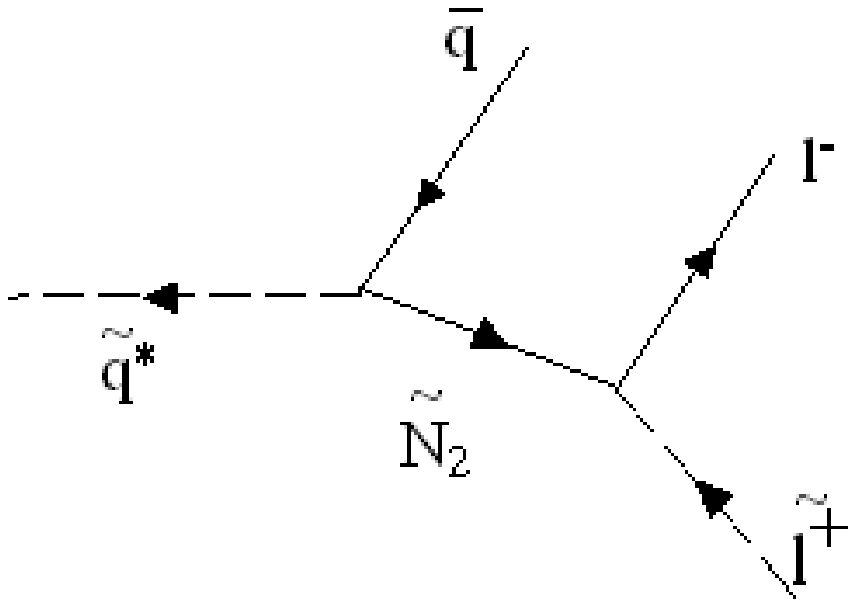} \hspace{.5 in}
\includegraphics[scale=0.5]{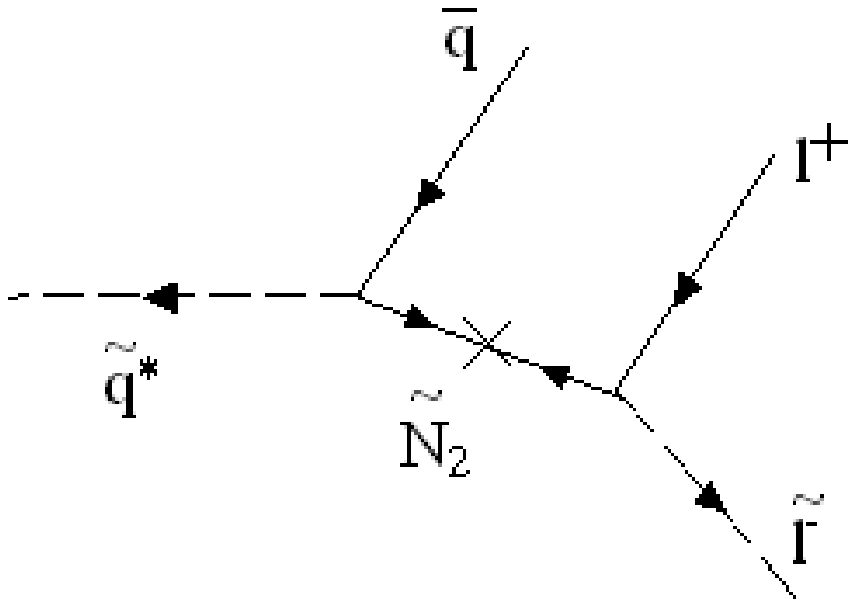}
\end{center}
\caption{The two conjugate modes starting from an anti-squark for the decay through a Neutralino.} \label{fig:
antiMajDecay}
\end{figure}

\subsection{Weak Decay with $q \bar{q}$ final states}

In principle, the decay into $q \bar{q}$ final states could contain similar spin correlations, since it just replace
the leptons in the second stage of the decay chains discussed in the previous section with quarks. However, in general
we can not determine the charge of the initial jet. Once we are forced to average over the two final states shown in
Fig.(\ref{fig: MajDecay}), all angular correlations are washed out.

On the other hand, if the decay products of the second decay are a third generation quark and quark partner we could,
in principle, recover some charge information. It will then be possible to extract some spin correlation from such
decay chains. The effectiveness of such decay channels require further careful studies taking into account the
efficiency of identifying charge of the third generation quarks.

\subsection{Weak Decay with $q W^{\pm}$ final state}

\begin{figure}[h]
\begin{center}
\includegraphics[scale=0.8]{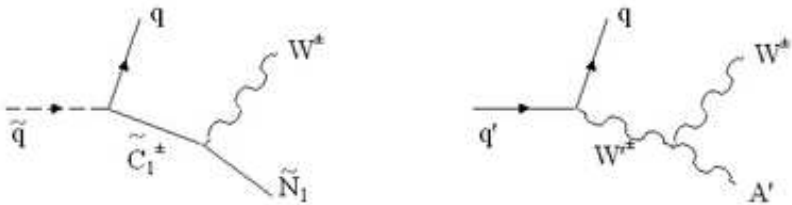}
\end{center}
\caption{The weak cascade decay of a quark partner through the non-Abelian vertex in supersymmetry (left) and Same-Spin
theories (right).} \label{fig: CASCweakBos}
\end{figure}

If the charged gauge boson partner is lighter than the leptonic partner then its decay into a $W^\pm$ and LSNP through
a non-Abelian vertex is usually the dominant decay mode. This channel is shown in Fig.(\ref{fig: CASCweakBos}) In the
supersymmetric case this coupling is at least partially chiral if $\tan\beta\ne 1$ and the higgsino is not considerably
heavier than the gauginos. If $m_{\tilde{q}}-m_{\tilde{C}}
>> m_q$, then the chargino is at least partially polarized (with respect to the axis defined by the
incoming squark and outgoing quark in its rest frame). In this case, since the chargino-neutralino-W coupling is also
in general chiral, correlations between the quark and the outgoing $W^\pm$ are present. The situation is a little more
subtle than that since the contributions from a cascade initiated by an up-type partner cancel those initiated by a
anti-down-type partner. However, due to the initial asymmetry between up quarks and anti-down quarks in the incoming
PDFs the signal is not washed out. This decay exhibits a linear dependence on the variable $t_{qW} = (p_q + p_W)^2$.

The corresponding distribution for the Same Spin scenario is very different. In the rest frame of the $W^{\prime}$ both
the incoming $q'$ and the outgoing $q$ are boosted and are mostly left-handed or mostly right handed. Hence the
$W^\prime$ is longitudinally polarized. As a result, this decay exhibits a quadratic dependence on $t_{qW}$  with a
positive coefficient.

We will present a detailed study of this channel in Section~\ref{sec: noleptons}. At this point, we just remark that
this is a more generic channel comparing with the channel requiring on-shell lepton partner in the decay chain. In
fact, the existence this decay channel is based on a very minimal set of assumptions about the spectrum, in which only
a heavy quark partner, a charged gauge boson partner and a LSNP are present. If the spectrum does not even allow for
this decay chain, we will not be able to extract any information from weak decays. This appears to be the most
promising channel.

\subsection{Weak Decay with $q Z$ final state}
There is a similar channel with a neutralino as the intermediate particle and a $Z^0$ in the final state (due to the
higgsino-higgsino-$Z^0$ coupling). This could be a potentially golden channel considering the leptonic decay of the
$Z^0$. Unfortunately, there are no angular correlations since the $\tilde{\bar{N}}_i\gamma_\mu P_\lambda \tilde{N}_j
Z^\mu$ vertex is not even partially chiral. The Same-Spin counterpart is slightly ambiguous. If the intermediate
particle is a heavy scalar partner of the higgs, there are no correlations. However, if the intermediate particle is
some heavy $Z'$ this might be the easiest channel to discover. As this is not a very generic case we will not pursue it
any further.

There is an additional complication concerning this process. When the $Z^0$ decays into quarks, this process is
experimentally indistinguishable from the previous one we consider involving a $W^\pm$. However, in most models it is
suppressed by a factor of $10-50$ with respect to the chargino channel owing to the higgsino origin of the coupling.
Therefore it does not present a serious background to it.

\subsection{Weak Decay with $qh$ final state}

\begin{figure}[h]
\begin{center}
\includegraphics[scale=0.8]{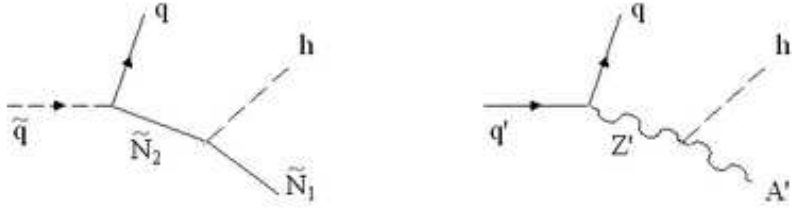}
\end{center}
\caption{The weak cascade decay of a quark partner through a heavy neutralino into a higgs and LSP, in supersymmetry
(left) and Same-Spin theories (right).} \label{fig: CASChiggs}
\end{figure}

The neutralino could also decay into a Higgs and LSP. This is shown in Fig.(\ref{fig: CASChiggs}). In the
supersymmetric case this process is possible because of mixing with the higgsino. Unfortunately, the
$h\bar{\tilde{N}}_1\tilde{N_2}$ vertex is not chiral and no correlation exists between the quark and higgs directions.

In the Same-Spin scenario this process is realized through the higgs coupling to the heavy gauge-bosons $g'v Z'_\mu
A'^\mu h$. In this case, a correlation between the higgs and outgoing quark exists and follow the same as those for a
massive gauge-boson decay into two bosons (the amplitude has a quadratic dependence on the variable $t_{qh}$ with a
positive coefficient).

This channel is quite generic and it is important to investigate it further. In certain cases, it might be possible to
replace the outgoing quark with an outgoing lepton (for example heavy slepton production as discussed below in
subsection \ref{subsec: leptonProd}). In a sense this is an orthogonal channel to that considered by Barr
\cite{Barr:2004ze} as it relies on the existence of heavy leptonic partners rather than light ones.

\subsection{Decay of Gluon partner}
In this section, we discuss the decay of the gluon partner into a quark and the quark partner. The quark partner
subsequently decays into another quark and missing energy. This is shown in Fig.(\ref{fig: CASCgluino}).  This would
certainly be the dominant channel of producing new physics particles if gluon partners are present in the spectrum.
This diagram might prove to be the dominant decay mode into missing energy.
\begin{figure}[h]
\begin{center}
\includegraphics[scale=0.8]{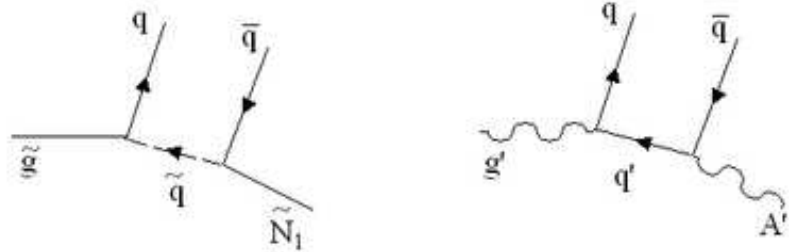}
\end{center}
\caption{The cascade decay of the gluon partner in supersymmetry (left) and Same-Spin theories (right).} \label{fig:
CASCgluino}
\end{figure}
Unfortunately, neither SUSY nor its Same-Spin counterpart have any spin effects present. The supersymmetric diagram
certainly does not involve any correlations between the two outgoing jets owing to the scalar nature of the
intermediate squark. In contrast the Same-Spin quark is indeed a fermion, however, its coupling to the gluon partner is
vector like. Therefore, it is unpolarized and its subsequent decay is isotropic. In the present work we will not
consider this channel any further, leaving a detailed study to a future publication.

If the spectrum is such that the gluon partner must decay into the LNSP via an off-shell quark the situation is quite
different. We discuss this issue further in subsection \ref{subsec: Offshell}.

\subsection{Strong Decay of Quark Partner}

\begin{figure}[h]
\begin{center}
\includegraphics[scale=0.8]{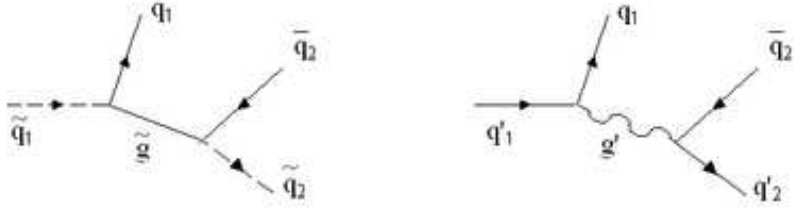}
\end{center}
\caption{The strong cascade decay of a heavy quark partner in supersymmetry (left) and Same-Spin theories (right).}
\label{fig: CASCsquark}
\end{figure}

Next we consider the strong decay of a quark partner. This scenario is slightly specialized as it relies on the
existence of a squark heavier than the gluino, but it is still generic enough to warrant consideration. The relevant
diagram is shown in Fig.(\ref{fig: CASCsquark}). If such a quark partner indeed exist this will be its dominant decay
mode.
The supersymmetric case still has no angular correlations between the outgoing jets owing to our experimental
limitations. As discussed above, the Majorana nature of the gluino makes it possible to observe correlations without
having chiral vertices. There are two diagrams, one with a mass insertion and the other without. The former involves
two outgoing quarks and the latter a quark and an antiquark. Unfortunately, all that we observe in the lab are two jets
and must average over the two contributions. Therefore the decay should have no dependence on the variable
$t_{q_{1}q_{2}} =(p_{q_1} + p_{q_2})^2$.

The Same-Spin case, however, does posses angular correlations between the outgoing jets and is distinguishable from the
supersymmetric one. As discussed above, the Same-Spin gluon is longitudinally polarized and we expect the decay to be a
quadratic function of the variable $t_{q_{1}q_{2}}$. The biggest challenge such a measurement faces is the signficant
background due to Standard Model processes. This may not be an insurmountable impasse. The subsequent decay of $q'$
together with hard cuts on missing energy might reduce the background dramatically. This is an important enough channel
with a clear enough signature (if isolated) to warrant further study which we hope to address in a future publication.

Notice that the second stage of the decay chain could involve third generation quarks and squarks. In this case, since
we might recover some charge information, this decay can be useful to determine the spin of the gluino. It is not
uncommon that the third generation squarks are lighter than the first two generation squarks. In particular, RGE
running and a large third generation Yukawa coupling usually results in a lighter third generation squarks. Therefore,
if $m_{\tilde{q}_3} \ge M_{\tilde{g}}$ or just slightly lighter, and $m_{\tilde{q}_{1,2}}>m_{\tilde{q}_3}$ there could
be a significant enhancement of branching ratio into third generation quark and quark partners. Even in the Same Spin
scenario, decaying into third generation quark and quark partner could help reduce the combinatorial background.

As mentioned above, in the case of left-right asymmetric spectrum, the two vertexes are effectively chiral and that
will effect the angular correlations.

\subsection{Decay from leptonic initial states}
\label{subsec: leptonProd}

Many of the channels discussed above, involved the weak decay of a quark partner. If lepton partners are heavy enough,
all such channels can be initiated by a lepton partner decay instead. The angular correlations are the same, only we
replace an outgoing jet with an outgoing lepton. Such a scenario can be realized through the Drell-Yan production of
heavy lepton partners.

Such a cascade has several advantages. First, jet combinatorics is not a problem. Second, we gain a lot more
information because charge and flavor is now available to us. This is extremely helpful. For example, in the weak decay
with $l,W^\pm$ final state, there is only one channel to consider and no averaging is needed.

On the other hand, a lepton partner at the beginning of a cascade is harder to come by. It could come from the decay of
heavy electroweak gauge boson partner, or a $Z'$ coupled to leptons. But, that is more model dependent. We also require
there to be several states below the lepton partner. This could sometimes require special arrangements. For example, in
the MSSM, we would require a mass hierarchy such as $M_2 > m_{\tilde{\ell}}>M_1> \mu$. This avenue looks promising in
certain regions of parameter space.

\subsection{Off-shell decays}
\label{subsec: Offshell}

So far, we have only considered on-shell decay processes. We saw that if the intermediate particle is a Dirac fermion
the interactions involved must be at least partially chiral. This conclusion is modified if that particle is off-shell.
Although on-shell decays usually dominate, there are special kinematical regions where we are forced to have off-shell
decays. For example:
\begin{enumerate}
\item If the quark partner is heavier than the gluon partner, gluon
  partner will be forced to decay through an off-shell quark partner.
\item The decay of a gauge boson partner to LNSP will be forced to go through an
off-shell W/Z and lepton/quark partners if the mass splitting is small. The virtual lepton-partner and virtual quark channels could be
particularly interesting since it brings in new spin information about the lepton partner. It could be important over a
large mass range of the lepton partner since the decay to off-shell W/Z is usually suppressed by mixing.
\end{enumerate}

To illustrate this point we consider the first example where the gluon partner decays through an off-shell quark
partner. The relevant diagrams are shown in Fig.(\ref{fig: CASCgluino}). The SUSY channel obviously has no correlations
since the squark is a scalar. However, in the Same-Spin scenario correlations are present. The amplitude
for this process is
\begin{equation}
\sum_{\rm pol}|\mathcal{M}|^2 \propto 2\frac{(m_{q'}^2 - q^2)(q^2- 2m_{g'}^2)(q^2-2m_{A'}^2)}{m_{g'}^2 m_{A'}^2} ~
t_{q\bar{q}} + f_0(q^2),
\end{equation}
where we neglected the trivial denominator. $f_0(q^2)$ is some complicated function of $q^2$, the momentum of the
internal quark partner, and the masses. It is irrelevant for this discussion. Notice that when the quark partner is
off-shell $q^2\ne m_{q'}^2$ and the coefficient of $t_{q\bar{q}} = (p_q + p_{\bar{q}})^2$ is non-zero. This linear
dependence of the cross-section on $t_{q\bar{q}}$ can in principle be distinguished from the SUSY case where there is
no dependence on $t_{q\bar{q}}$. Further study is needed to explore the observability of this effect in different
models.

Similar considerations apply for the other case of a gauge-boson partner decay to LNSP via an off-shell slepton.

\section{Determining spin without leptonic partners}
\label{sec: noleptons}

In this section we will explore the decay of a quark partner into a charged weak partner which consequently decays into
a $W^\pm$ and missing energy. Let's begin with the supersymmetric case. The squark-quark-chargino vertex is given by
\cite{Chung:2003fi} (we are ignoring the CKM and super-CKM matrices as they are quite irrelevant to the following
discussion),
\begin{equation}
\label{eqn: sqCqCoup}
 \mathcal{L}_{q\tilde{q}\tilde{C}^+} = -g_2\left( \bar{u}P_R (U_{11}\tilde{C}_1 +
U_{21}\tilde{C}_2) \tilde{d} + \bar{d} P_R (V_{11}\tilde{C}^c_1 + V_{21}\tilde{C}^c_2) \tilde{u}\right)
\end{equation}
where $U_{ij},V_{ij}$ are the matrices diagonalizing the Chargino's' mass matrix. We are assuming that the chargino is
dominantly a gaugino and ignore the direct quark-squark-higgsino couplings. The more important vertex is the
chargino-$W^+$-neutralino coupling,
\begin{equation}
\mathcal{L}_{W^-\tilde{C}\tilde{N}} = g_2 W_\mu^- \bar{\tilde{N}}_i\gamma^\mu\left(O_{ij}^L P_L +O_{ij}^R P_R\right)
\tilde{C}_j
\end{equation}
where,
\begin{eqnarray}
\label{eqn: CNWcoup}
 O_{ij}^L &=& -\frac{1}{\sqrt{2}} N_{i4} V_{j2}^* + N_{i2}V_{j1}^*\\\nonumber
 O_{ij}^R &=& \frac{1}{\sqrt{2}} N_{i3}^*
U_{j2} + N_{i2}^*U_{j1}
\end{eqnarray}
and $N_{ij}$ are the mixing matrices for the Neutralino. This interaction is usually at least partially chiral (when
$tan\beta \ne 1$). Therefore we expect the amplitude to have some $t_{qW}=(p_q+p_W)^2$ dependence, with a coefficient
given by the difference of the couplings in equation (\ref{eqn: CNWcoup}). Indeed, in the narrow width approximation
$q^2 \rightarrow m^2_{\tilde{C}}$ we get,
\begin{equation}
\label{eqn:SUSYsignal1}
 |\mathcal{M}|^2 \propto \frac{1}{2}\left(\frac{m_{\tilde{C}}^2(m_{\tilde{C}}^2 - m_{\tilde{N}}^2 -
  2m_W^2)} {m_W^2}\right)~(a_L^2-a_R^2) ~t_{qW} + f_0(q^2,m_i)
\end{equation}
where $f_0(S,m_i)$ is a polynomial given in the appendix. The 3-body phase-space differential volume can be written in
terms of $q^2$ and $t_{qW}$ (see for example \cite{Barger:1987nn}),
\begin{equation}
  dPS_3 = \frac{1}{128\pi^3 m_1^2} dq^2 dt_{qW}
\end{equation}
with appropriate kinematic boundaries. In the narrow-width approximation the integration over $q^2$ is trivial and
simply removes the denominator in equation (\ref{eqn:SUSYsignal1}) and replaces $q^2\rightarrow m_{\tilde{C}}^2$.
Therefore, the angular correlations in this decay depend on the difference $(a_L^2-a_R^2)$.

\begin{figure}
\begin{center}
\includegraphics[scale=0.3,angle=270]{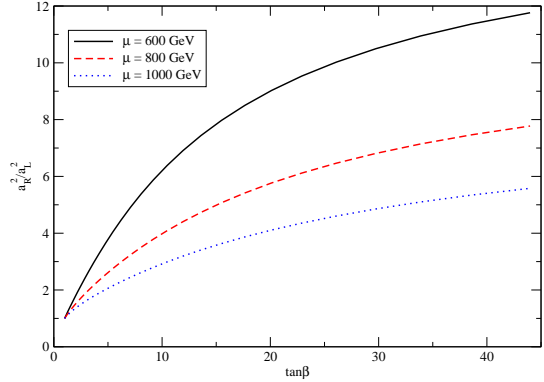}
\end{center}
\caption{The ratio of the right to left couplings $a_R^2/a_L^2$ as a function of $\tan\beta$ for three different values
of the $\mu$-parameter. $M_1=100\GeV$ and $M_2=500\GeV$ are fixed.} \label{fig: TanbetaScan}
\end{figure}

In Fig.(\ref{fig: TanbetaScan}) we plot the ratio $a_R^2/a_L^2$ as a function of $\tan\beta$ for a few values of
$\mu$-parameter with $M_1 = 100\GeV$ and $M_2 = 500\GeV$. This ratio is quite different than unity for most choices of
the parameters.

\begin{figure}[h]
\begin{center}
\includegraphics[scale=.3,angle=270]{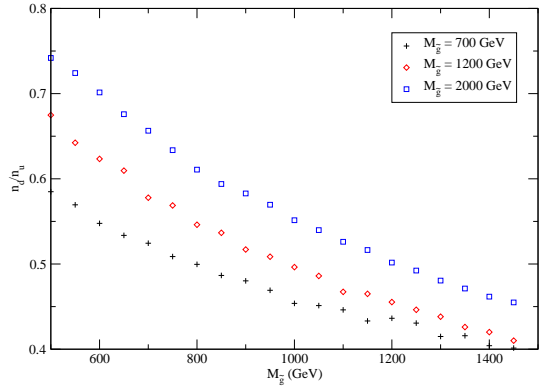}
\end{center}
\caption{The ratio of down squarks to up squarks in production as a function of their mass for different gluino masses,
$m_{\tilde{g}} = 700,~1200,~2000\GeV$ (black-solid, red-dashed, green-points).} \label{fig: DUratio}
\end{figure}

There is one additional complication in the SUSY case. As seen from equation (\ref{eqn: sqCqCoup}) there are two
contributions to any process involving the chargino. One comes from the coupling to the up squark, while the other
comes from the coupling to the anti-down squark. Therefore, as shown in details in the appendix these two contributions
differ by the sign of the coefficient of $t_{qW}$. Since we cannot distinguish between a jet coming from a down quark
and that coming from an anti-up quark, we must average over the two contributions. Fortunately, due to the composition
of the proton there are more up-like squarks produced than down-like squarks. Their ratio in production is shown in
Fig.(\ref{fig: DUratio}) as a function of their mass and gluino mass.

\begin{figure}
\begin{center}
\includegraphics[scale=0.25,angle=270]{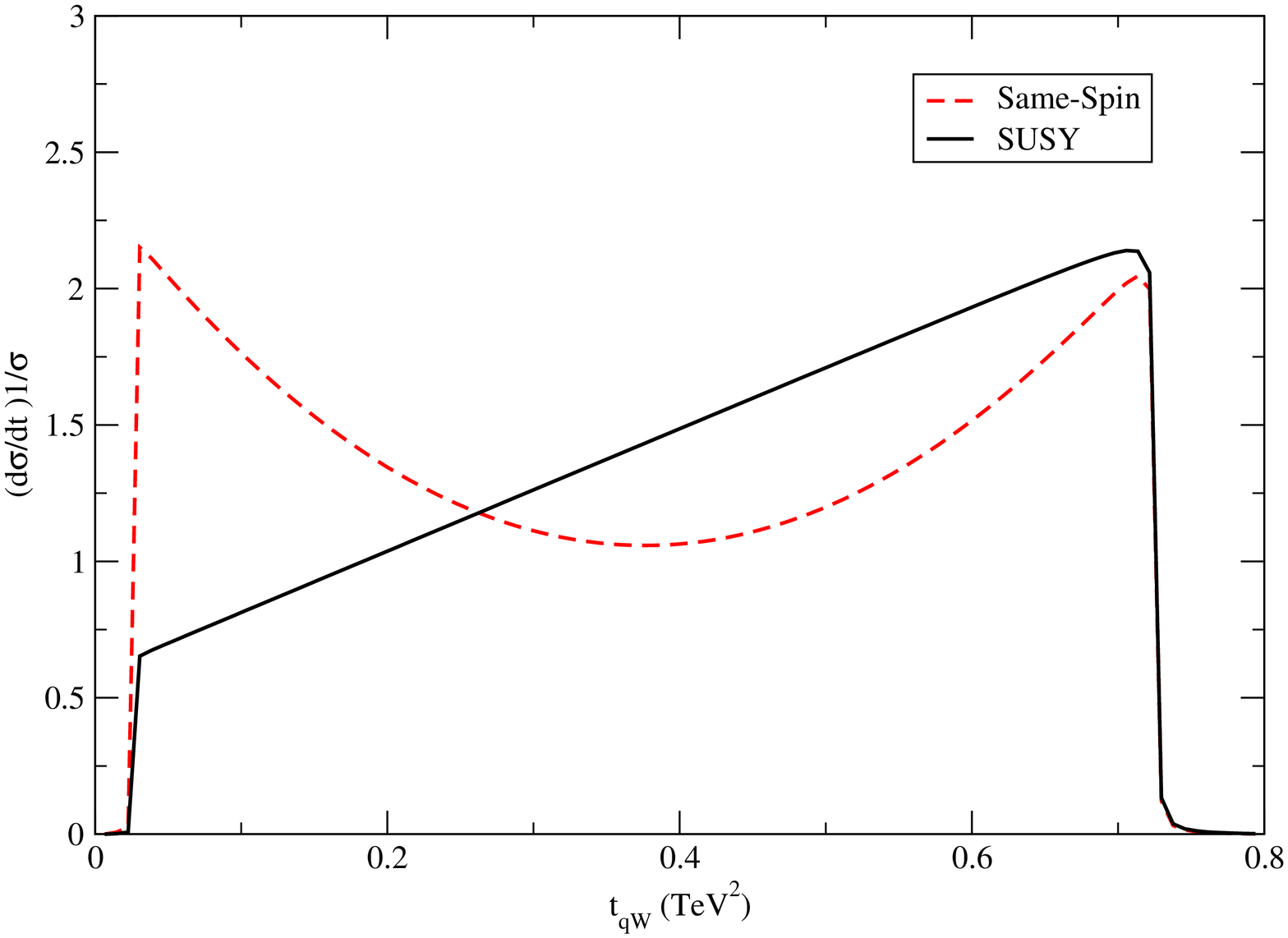}
\includegraphics[scale=0.25,angle=270]{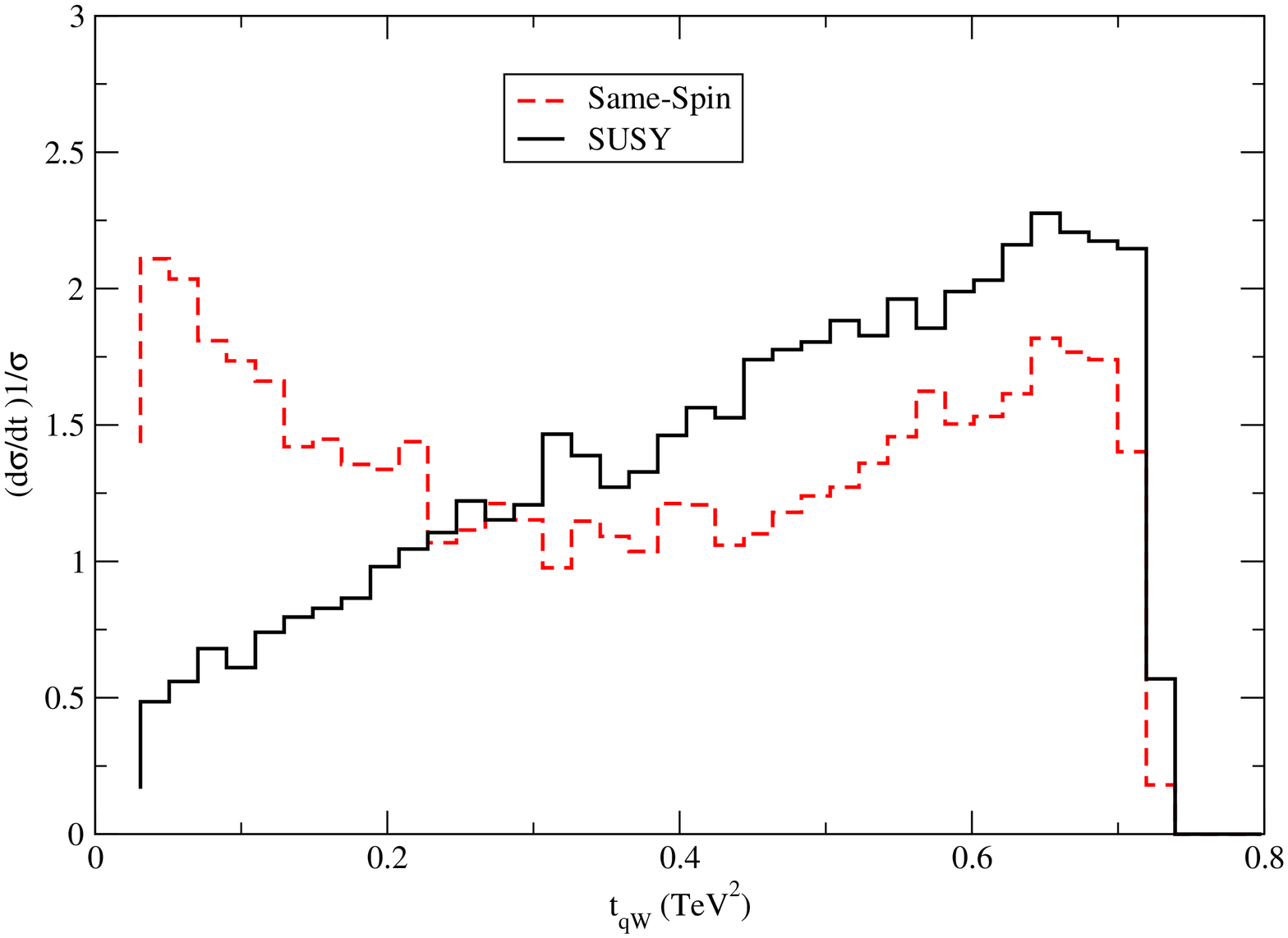}
\end{center}
\caption{Theoretical curves (Left graph) for the q-W correlations in the two models (solid,black - SUSY, dashed,red -
Same-Spin). Cross-section is plotted against the $t_{ql}$ variable. The right plot shows the Monte-Carlo simulation.
Both graphs are normalized to unit area.} \label{fig: thrcurvesQW}
\end{figure}

Let us contrast this with the corresponding process in Same-Spin theories. The intermediate particle is a spin-1
Same-Spin $W^\pm$. As we argued before in equation (\ref{eqn: Wqqkk coupling}) it is dominated by the longitudinal
mode. Therefore  its subsequent decay is dominated by the angular distribution of the second equation in
(\ref{eqn:MgbPolBos}). Therefore the $W^+$ boson is preferentially collinear or anti-collinear with the jet. We expect
$a_2 (t_{qW})^2 + a_1 t_{qW}+ a_0$ dependence with $a_2> 0$ and $a_1 \le 0$. The computation is quite involved, but the
final expression is indeed,
\begin{equation}
\label{eqn: KKsignal1}
 |\mathcal{M}|^2 = \frac{1}{(q^2-M^2)^2}\left(F_0(q^2,m_i) + F_1(q^2,m_i) t_{qW} +F_2(q^2,m_i)(t_{qW})^2 \right)
\end{equation}
where the $F_i$'s are given in the appendix. It is not hard to show that $F_2(M^2,m_i) > 0$ and $F_1(M^2,m_i) <0$. The
shape of the resulting cross-section is plotted in Fig.(\ref{fig: thrcurvesQW}) against the corresponding SUSY
cross-section. The behavior for small $t_{qW}$ is distinctly different than the supersymmetric case. The reason is
clear. Small values of $t_{qW}$ corresponds to $W^+$ being collinear with the jet, which is forbidden in the
supersymmetric process, but preferred in the Same-Spin case. We have also included in Fig.(\ref{fig: thrcurvesQW}) the
results of the Monte-Carlo simulation.

\subsection{Experimental Observable - lepton-jet correlation}
Of course, we have to take into account  the fact that  $W^+$ cannot be observed directly and only its decay products
can be measured. If it decays to quarks and it is possible to distinguish these two jets from the rest by
reconstructing the $W^+$ (as it is on-shell) the signal might still be very strong. We include standard cuts to reduce
Standard Model background. We also consider the contribution from other new physics processes with identical final
states. We argue that this signal is still a strong candidate for spin determination\footnote{A detailed study of
Standard Model background with sophisticated jet analysis is beyond the scope of this paper. We will begin to address
this issue in future publications.}. In this subsection, however, we concentrate on the other option, namely, the
leptonic decay of the $W^\pm$. The main challenge we face is that we cannot reconstruct the $W^+$ as the neutrino is
unobservable. To investigate the resulting signal we used the Monte-Carlo event generator HERWIG
\cite{Corcella:2002jc}. We implemented spin-correlations for massive spin-1 particles and the details can be found in
appendix \ref{app: HERWIG}. For the Same-Spin production matrix elements we used the ones quoted in
\cite{Smillie:2005ar} \footnote{Except for the matrix element $\mathcal{M} (qq \rightarrow q_1^*q_1^* )$ which was
taken from \cite{Macesanu:2002hg} since the one quoted in \cite{Smillie:2005ar} seemed to give production rates which
are too large. Strictly speaking, those are inappropriate for the model we consider (as they assume degenerate spectrum
and only one mass scale, namely $1/R$, the compactification radius). However, nothing in our discussion seems to relay
heavily on the precise production cross-section and we do not expect any major modification to the conclusions below.
To simplify the analysis we switched off cluster formation, heavy hadron decays and the underlying soft event.}

\begin{figure}[h]
\begin{center}
\includegraphics[scale=0.3,angle=270]{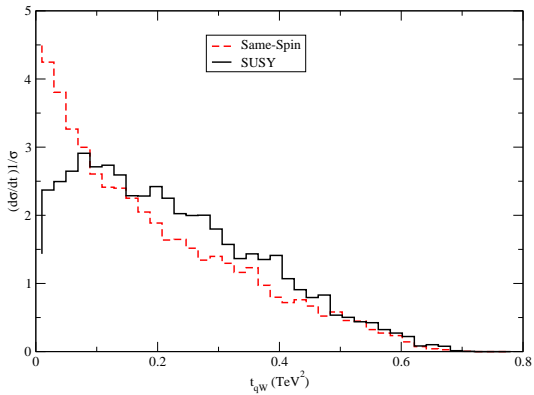}
\end{center}
\caption{Monte-Carlo simulation for the quark-lepton correlations in the two models (solid,black - SUSY, dashed,red -
Same-Spin).  Normalized cross-section is plotted against the $t_{ql}$ variable. The data sets contain $\sim 13,000$
events each.} \label{fig: QLMCcurves}
\end{figure}

We expect much of the difference to be washed out once the $W^+$ is allowed to decay into leptons. The only observables
we are left with are the momenta of the outgoing jet and that of the lepton. Therefore, we will plot the cross-section
as a function of the invariant mass $t_{ql} = (p_q + p_l)^2$. Fig.(\ref{fig: QLMCcurves}) shows the Monte-Carlo
simulation results when the lepton is correlated with the jet from it own branch. In practice we have no why to tell
which jet came from which branch and we amend this below.

While the behavior is indistinguishable at high $t_{ql}$, it is certainly very different at low $t_{ql}$. We can
attribute this to the fact that in the SUSY model, the $W^+$ is very unlikely to be collinear with the jet, whereas the
opposite is true for the Same-Spin model.

\begin{figure}[h]
\begin{center}
\includegraphics[scale=.25,angle=270]{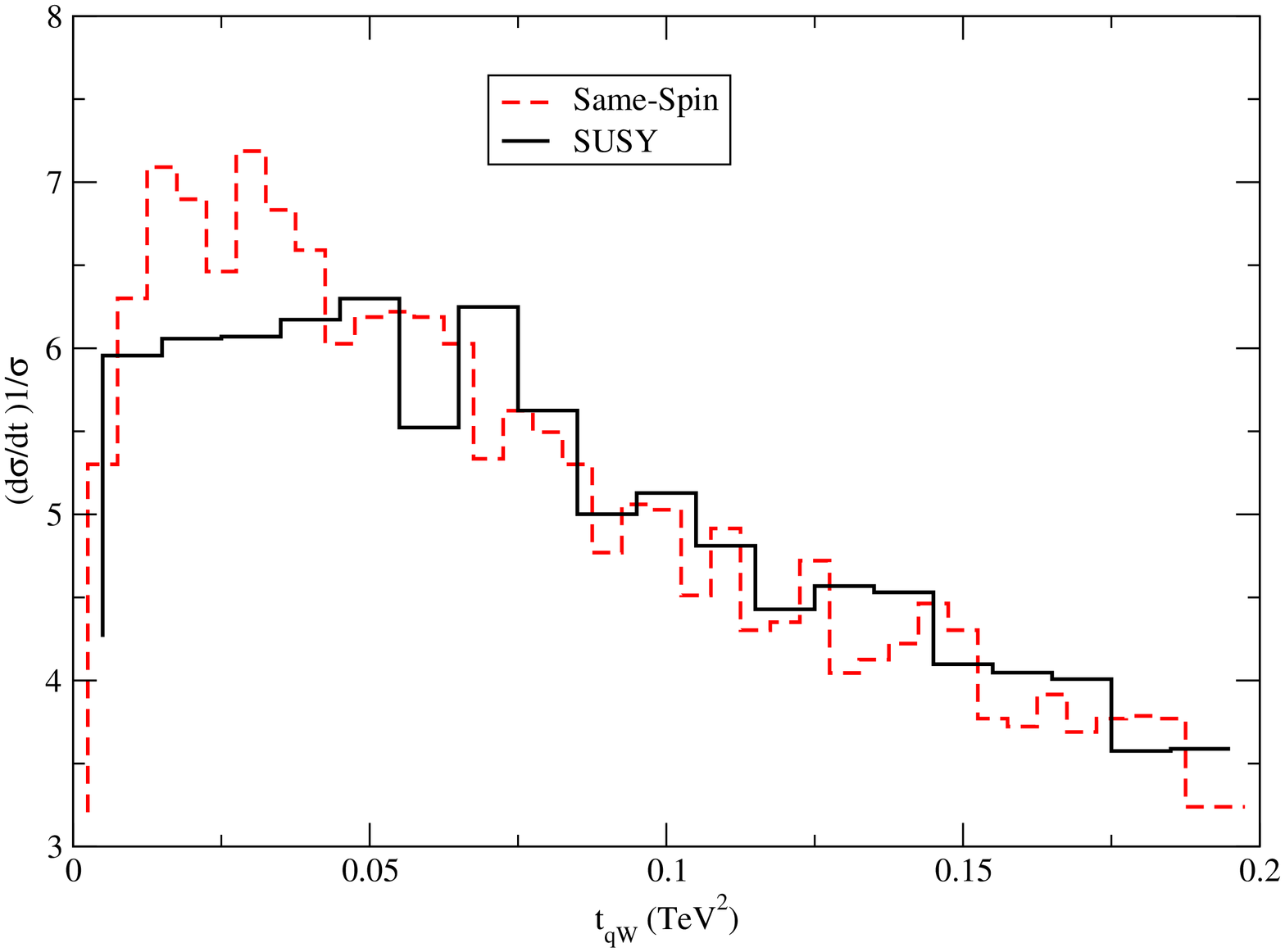}
\includegraphics[scale=.25,angle=270]{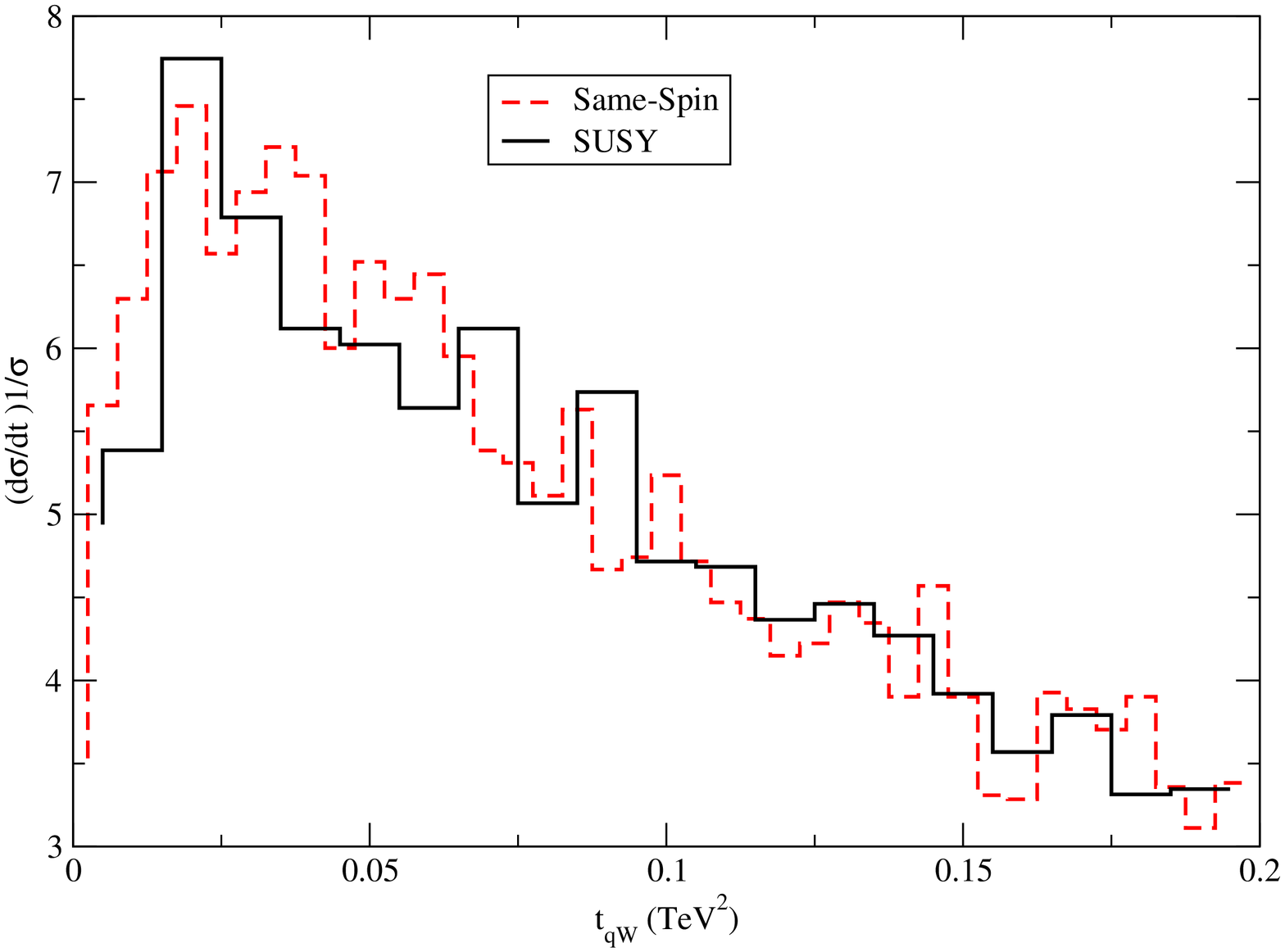}
\end{center}
\caption{Monte-Carlo simulation for the jet-lepton correlations in the two models (solid,black - SUSY, dashed,red -
Same-Spin) with no knowledge of the correct pairing. Normalized cross-section is plotted against the $t_{ql}$ variable.
The graph on the left is for 2 jets and 1 lepton events. The graph on the right corresponds to 2 jets 2 opposite sign
leptons events. The data sets contain $\sim~17,000$ events with 1 lepton and $\sim~9,000$ events with 2 leptons for
each of the models.} \label{fig: QL12MC}
\end{figure}

This plot suggests that the two models are still distinguishable from one another even if the $W$ cannot be fully
reconstructed. When taking into account the jet combinatorics by pairing the lepton with both jets in the event the
results do alter. In Fig.(\ref{fig: QL12MC}) we plot the cross-section against $t_{ql}$ for events with two jets and
one or two leptons (the two-leptons case corresponds to both branches decaying into a lepton). The single lepton
diagram still exhibits the flattening of the cross-section in the SUSY case for low $t_{ql}$. This is in contrast with
the rising cross-section for the Same-Spin model. It is possible that close analysis of low $t_{ql}$ can pick up this
difference once real data is available. There is no obviously observable spin dependence in the second case with two
leptons in the final state.

\subsection{Experimental Observable - jet-$W$ correlation}

\begin{figure}[h]
\begin{center}
\includegraphics[scale=.5]{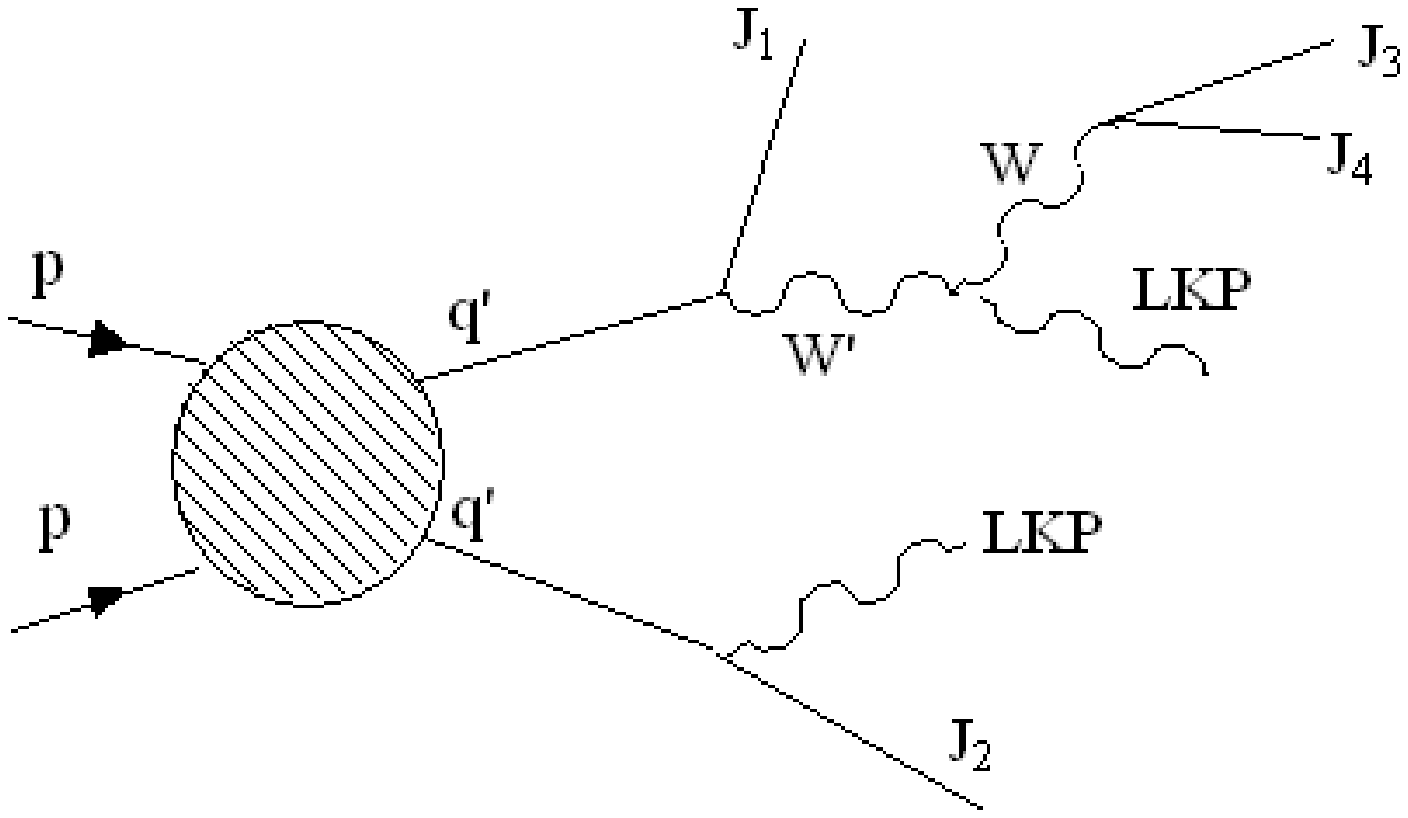}
\includegraphics[scale=.5]{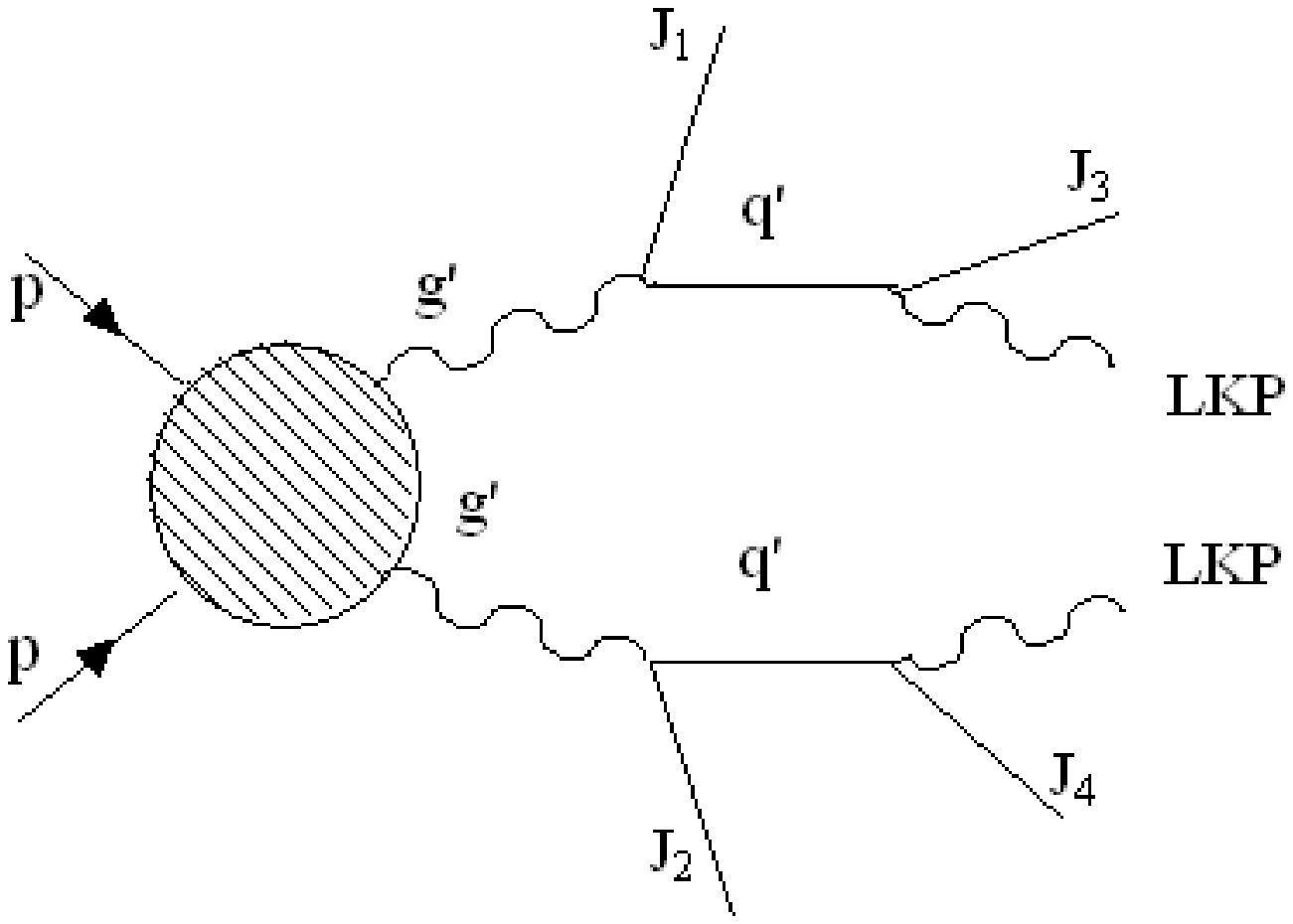}
\end{center}
\caption{4-jet event topology for the Same-Spin theory. The diagram we are interested in (left) together with a
possible Same-Spin irreducible background (right)} \label{fig: 4jEvent}
\end{figure}

In this section we take up the second possibility, namely that the $W^\pm$ decays into two jets. The advantage is the
ability of fully reconstructing the four momenta of the $W^\pm$'s. There are disadvantages as well. First and foremost,
very naively the Standard Model background is significant. Second, since jets are involved, momentum determination
involves some amount of smearing. This would affect the reconstruction of $W^\pm$ as well as the angular correlations.
Third, it is very hard to distinguish between a $W^\pm$ and a $Z^0$ and so when investigating background we must
consider both. We do not attempt a full Standard Model background analysis, however, there are several reasons why it
might be possible to reduce such a background. First, hard cuts on missing energy can yield a fairly clean sample of
beyond the standard model physics. Second, we can easily have additional leptons in the process we consider. Simply let
the quark partner in the other branch (shown in Fig. (\ref{fig: 4jEvent})) to decay into a $Z^{\prime}$ or
$W^{\prime}$. It is not hard to imagine other possible channels for producing leptons in the final state. The problem
of background reduction is beyond the scope of this paper and we will assume that some non-negligible set of events,
containing new physics, can be isolated and analyzed.

\begin{figure}[h]
\begin{center}
\includegraphics[scale=.4,angle=270]{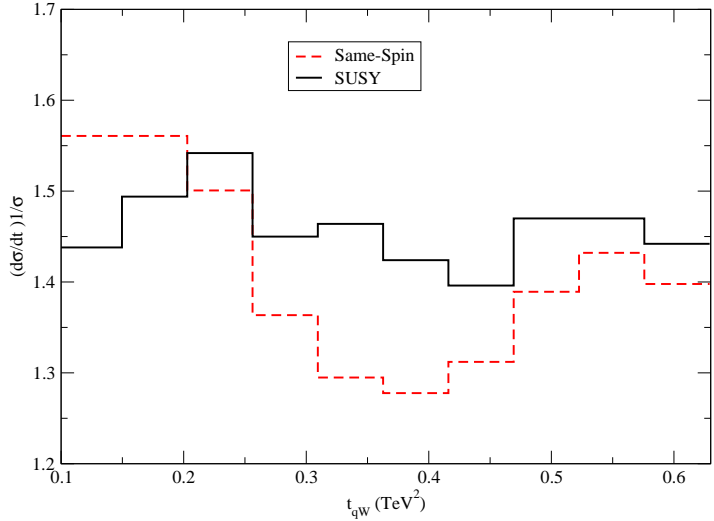}
\end{center}
\caption{Considering only 4-jet events this is a histogram of the invariant mass $t_{W,j_l}$ for each of the models:
SUSY (solid-black), Same-Spin (dashed-red). The histogram is normalized to unit area. The two data sets contained $\sim
9,000$ and $2,000$ events, respectively. The normalized error is approximately $\sqrt{N}\sim 0.04$.} \label{fig:
JetsWmc}
\end{figure}

We consider 4-jet events with a typical event topology shown in Fig.(\ref{fig: 4jEvent}) together with a possible
background from another new physics process. In every event we try to reconstruct the $W^\pm$ from two of the 4 jets
and then form the invariant mass $t_{qW}$ with the two remaining jets. If more than one pairing reconstructs $W$ it is
regarded as failure and the event is discarded. Our cuts involve $\slashed{P}_T > 200\GeV$ and $\eta <4.0$. We make no
attempt in trying to order the jets by the magnitude of their transverse momentum or some more sophisticated ordering
(This very naive approach yields a significant difference between the two models as shown in Fig.(\ref{fig: JetsWmc})).
There can certainly be potential improvements on this measurement by using more kinematical information of the jets.
The linear behavior vs. the the quadratic behavior is still visible on top of the background coming from the wrong
pairing.

\begin{figure}[h]
\begin{center}
\includegraphics[scale=.5]{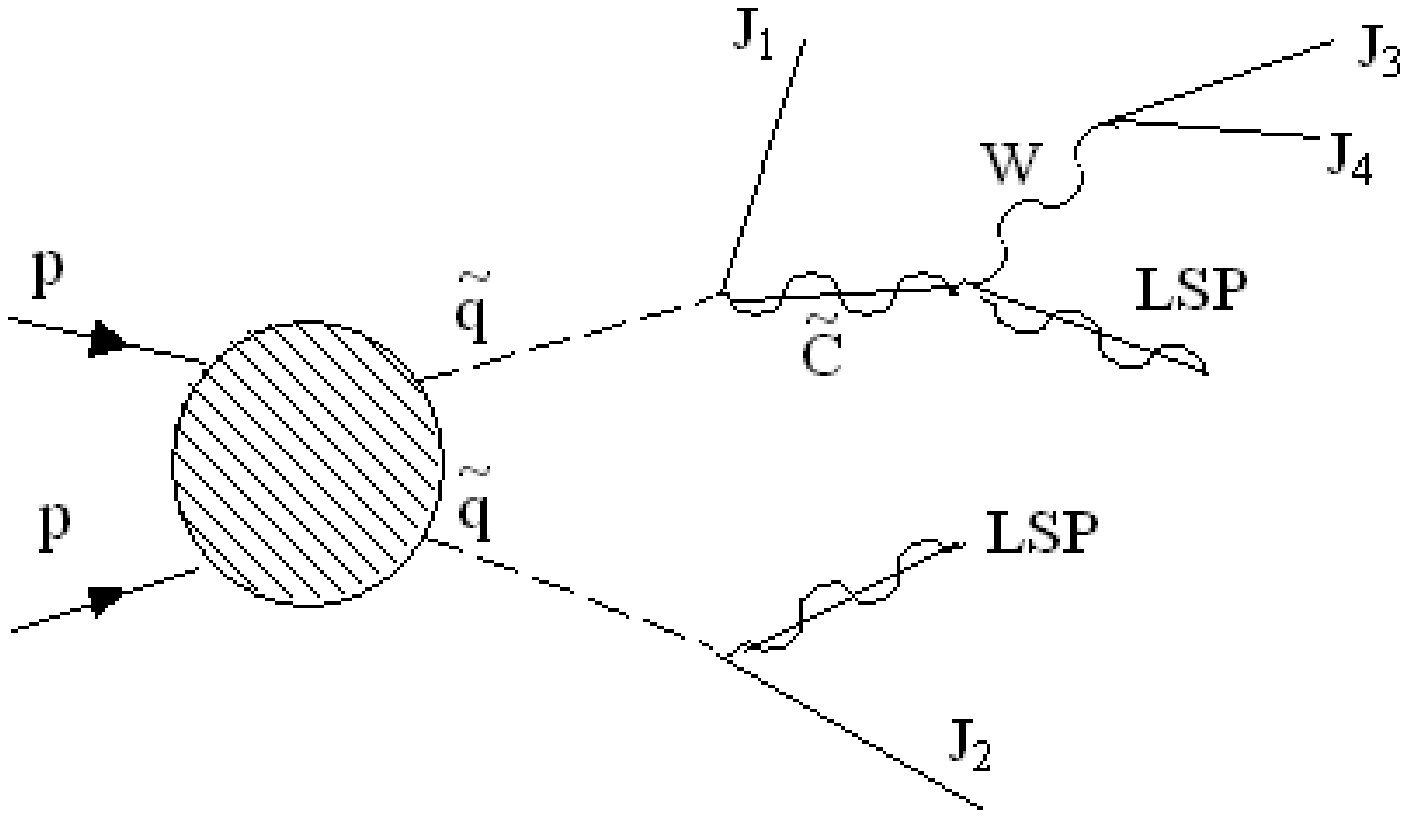}
\includegraphics[scale=.5]{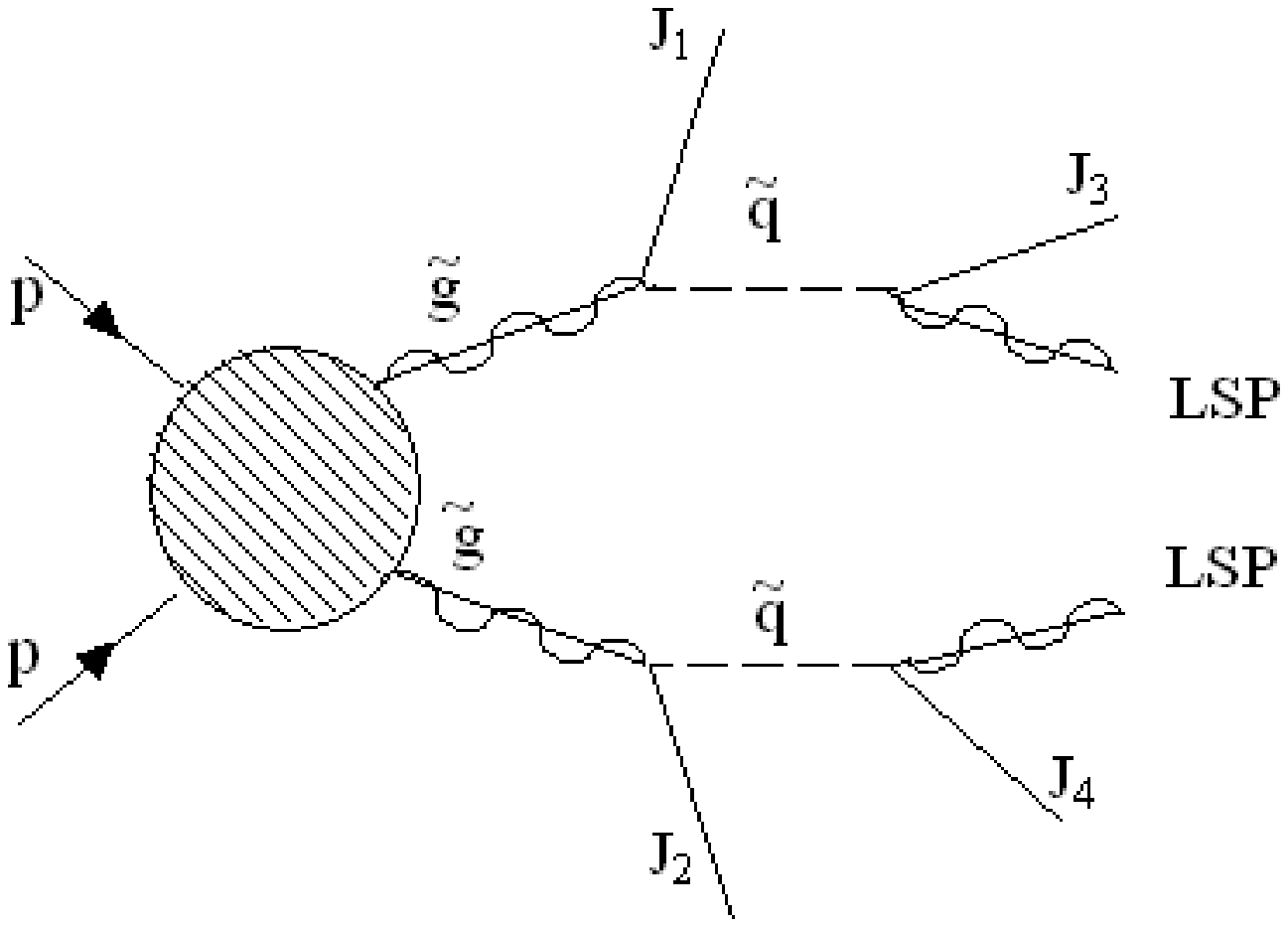}
\end{center}
\caption{4-jet event topology for SUSY. The diagram we are interested in (left) together with a possible SUSY
irreducible background (right)} \label{fig: 4jEventSUSY}
\end{figure}

\begin{figure}[h]
\begin{center}
\includegraphics[scale=.3,angle=270]{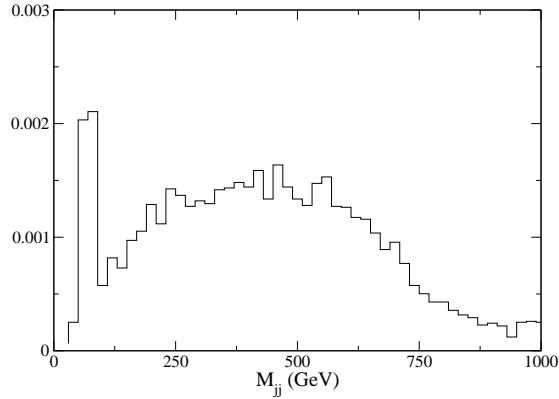}
\end{center}
\caption{Considering only 4-jet events we plot a histogram of the invariant mass $m_{j_k,j_l}^2 = (p_k+p_l)^2$ of all
possible pairs of jets. The graph is normalized to unit area.} \label{fig: qWBG}
\end{figure}

There could also be background from different processes in the same model of new physics with identical final states.
One such channel is shown on the right in Fig. (\ref{fig: 4jEvent}) for Same-Spin theory and Fig. (\ref{fig:
4jEventSUSY}) for the SUSY case. We consider all events with 4-jets. Then we construct the invariant mass
$m_{j_k,j_l}^2 = (p_k+p_l)^2$ for every possible pair. We used HERWIG internal algorithm for jet progenitor formation\footnote{This object is not a jet, but would become one after a proper experimental jet algorithm is applied.}. We include
jet smearing effects \cite{Han:private} using the ATLAS specs \cite{unknown:1999fr}. The results are shown in
Fig.(\ref{fig: qWBG}). The irreducible background does not seem to pose a serious concern as the $W$ peak is clearly visible. This should come as no surprise. With a random pairing of such energetic jets, the chance of
reconstructing a quantity with $< 100\GeV$ is fairly low. This can probably be made even sharper with some simple cuts
on jet energy. For example, the initial jets from the squark decay tend to be more energetic than those coming from the
$W$.

The above discussion is by no means a proof that the reconstruction of the $W$ is an easy task. It merely serves to show that it is not hopeless and to encourage further study of this possibility involving a proper detector simulator and a more detailed analysis of the background.

\subsection{Scanning $M_1$ and $M_2$}

It is important to map out the regions of parameter space in which the different channels are useful. In this section
we present the results of a scan covering the subspace spanned by $(M_1,M_2)$. We also set $m_{W'}=M_2$ and
$m_{A'}=M_1$. While there are other parameters which effect the results (such as $\tan\beta$, $m_{\tilde{q}}$, $\mu$,
etc.) we focus on those two parameters, $(M_1,M_2)$, which are the most determinantal to the observability of angular
correlations for the channel we consider.

When performing such a scan, we wish to assign a number to quantify the difference between, say, the distribution
describing the SUSY scenario and that of the Same-Spin scenario. There is no unique choice for such a number. Moreover,
such an assignment can be problematic as it can overlook differences that a more careful analysis would pick up.
Therefore, the results of this section should be understood with the following proviso in mind: the numbers assigned
for the different points in parameter space carry only relative importance among themselves and have very little
absolute meaning. They indicate that in certain regions spin determination is easier as compared with other regions.

There is one more point to keep in mind. We will compare the distributions produced for the two models by matching
their spectrum. This is incorrect. One should match the cross-section first as that is the actual experimental
observable. Unfortunately, we had poor control over the production rates for the Same-Spin scenario\footnote{The
production cross-sections given in \cite{Smillie:2005ar} for the Same-Spin theory have only one adjustable parameter,
namely the inverse compactification radius.} However, this is not entirely misleading. As part of our results we will
consider the observability of each of the models against phase-space. Therefore, if nothing else, we are able to tell
when spin effects are present at all.

To quantify the difference between the distributions we use the Kolmogorov-Smirnov test for goodness-of-fit. In this
test, the cumulative distribution functions (CDF) of both data sets are compared. The D-statistics is simply the
maximum vertical difference between the CDF's. In other words,
\begin{equation}
  D = sup\left|F_1(x)-F_2(x)\right|
\end{equation}
where $F_1(x)$ and $F_2(x)$ are the CDF's for the two data sets. The p-value assigned to the D-statistics is low when
the two data sets come from different underlying distributions. There are several advantages to the Kolmogorov-Smirnov
test. First, it is a non-parametric test so it applies to general distributions. Second, it is independent of the way
we choose to histogram the data.

\begin{figure}[h]
\begin{center}
\includegraphics[scale=.5,angle=270]{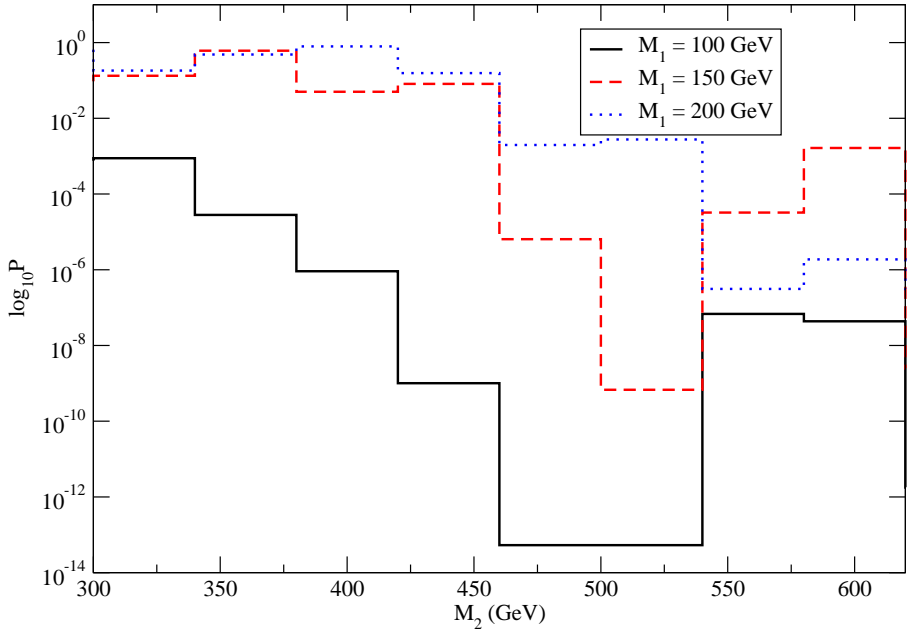}
\end{center}
\caption{A plot of the p-value vs $M_2$ for different values of $M_1$. We set $m_{W'}=M_2$ and $m_{A'}=M_1$. The
p-value is lower when the two data sets are more distinguishable. The other parameters were fixed at $m_{\tilde{q},q'}
=1000\GeV$,$ m_{\tilde{g},g'} =1200\GeV$. We set $\tan\beta = 10$ in the supersymmetric case. Both data sets contained
$\sim 10,000$ events. Notice that when the mass splitting between $M_1$ and $M_2$ is small, i.e. the left-hand side of
the plot, it is harder to distinguish these two scenarios. Supersymmetry and Same-Spin scenario become more
distinguishable as the mass splitting increases. However, as $M_2$ approaches $m_{\tilde{q},q'}$ the difference is
again diminished.} \label{fig: scan100K}
\end{figure}

In Fig. (\ref{fig: scan100K}) we plot the p-value as a function of $M_2$ for three different values of $M_1$. The quark
partner mass is $m_{\tilde{q},q'} =1000\GeV$, the gluon partner mass is $ m_{\tilde{g},g'} =1200\GeV$. We set
$\tan\beta = 10$ in the supersymmetric case. For every value of the parameters we produced $100,000$ events out of
which only about $10\%$ passed the different cuts.

The scan matches our expectations. When $M_2 \rightarrow M_1$ the LSP and the $W$ are produced at rest and there the
correlations with the polarization axis defined by the quark partner are diminished. Also, when $M_2\rightarrow
m_{\tilde{q},q'}$ it becomes harder to distinguish the two data sets. This is also expected because in the Same-Spin
case the quark partner, $q'$, is not at all boosted in the rest frame of the $W'$. Therefore, $W'$ is not polarized.

The results are encouraging. While a more sophisticated analysis (better cuts, better fitting, more realistic collider
simulation, etc.) is certainly warranted, these initial results seem to indicate that spin determination is possible.
As pointed out before, one should not confine the analysis to special benchmark points, but rather attempt a scan over
the parameter space. In this way, an inclusive strategy, combining several channels, can be devised to cover different
exclusive regions of the parameter space.

\section{Conclusions and Future Directions}
\label{sec: Future}

We have systematically studied the possibility of measuring the spin of new particles in a variety of cascade decays.
Generally, the existence of a LNSP renders the reconstruction of the momenta of new physics particles impossible.
Therefore, we focused on distributions of relativistically invariant variables. We identified a set of decay channels
which are useful for spin determination.

A general lesson of this study is that even though spin correlations are present in a variety of decay channels, the
viability of any particular channel is always confined to certain kinematical regions. As a result, different models of
new physics with different spectra tend to give very different useful channels for spin determination. Different
strategies will have to be employed at the LHC. Therefore, we emphasize that it is important not to confine ourselves
to any particular benchmark model or any particular decay mode. We should instead explore the effective range of all
possible decay channels and devise techniques for using them efficiently. In this sense, our survey of potentially
useful decay channels is the initial step of an important task for which many detailed studies still need to be done.

As part of this program we studied the decay channel $\tilde{q}\rightarrow q + \tilde{C}^\pm \rightarrow q + W^\pm +
LSP$. A scan over the chargino and neutralino masses was performed. We found that as long as the spectrum is not too
degenerate the prospects for spin determination are rather good.

One of the most important challenges is to measure the spin of gluon partners. The difficulty in such a study is that
the decay products carrying the spin information of the gluon partner are usually jets, missing the charge information.
In addition, such channels usually have larger combinatorial background and Standard Model background. A similarly
challenging task is a direct measurement of the spin of the quark partner.

 We have considered only kinematical variables constructed out of two objects of the decay
products. It is in principle interesting to study the possibility of  using more complicated kinematical variables.

Notice that decay channels involving leptons are generally more promising than those involving only jets. This result
stems from the fact that we have charge and flavor information from the lepton, which could help us in separating the
channels. We have assumed no such information from jets. Therefore, we have limited information from channels decaying
into quarks, except for the third generation. Moreover, losing charge and flavor information from the jet significantly
increases the combinatorial background, since typical new physics signals for the scenarios considered in this paper
almost always contain several hard jets. Therefore, any potential information about the charge and flavor of the
initial parton of the jet will be very helpful in spin determination.

We would like to remark that reducing combinatorial background is very important in extracting more information about
the underlaying new physics. This lead the authors of \cite{Arkani-Hamed:2005px} to conclude that to what extent jet
charge could be measured is an important study for LHC experiments.

We have not studied the possibility of measuring spin in the production. This is certainly a very important area to be
investigated carefully. Barr \cite{Barr:2005dz} has studied the measurement of the spin of muon partners using angular
distribution from pair production. In principle, angular distributions of production of other partners should carry
similar information. This is a subject currently under investigation. One of the main complications is the existence of
t-channel productions. Such production channels bring in more partial-waves and tend to wash out characteristic angular
distributions.

We would also like to emphasize that although we have only compared SUSY with the Same-spin scenario, the general rules
we have developed in the paper are easily applicable to any generic scenario with different spin content of new physics
particles.


\textbf{Acknowledgments}: We would like to thank the following people for useful discussions: Nima Arkani-Hamed,
Michael Peskin, Gui-Yu Huang, Jesse Thaler and especially Tao Han whose encouragement and suggestions were very
helpful.

\appendix
\renewcommand{\theequation}{A-\arabic{equation}}
\setcounter{equation}{0}
\section{Matrix elements calculations}
\label{app: MEs}

In this appendix we present the explicit matrix elements for the decays considered in the text. We begin with the
Same-Spin decay channel $q^{\prime} \rightarrow q + W^{\prime} \rightarrow q + W + A'$. The matrix element is given by,
\begin{eqnarray*}
  i\mathcal{M} &=& \vcenter{
\includegraphics[scale=0.5]{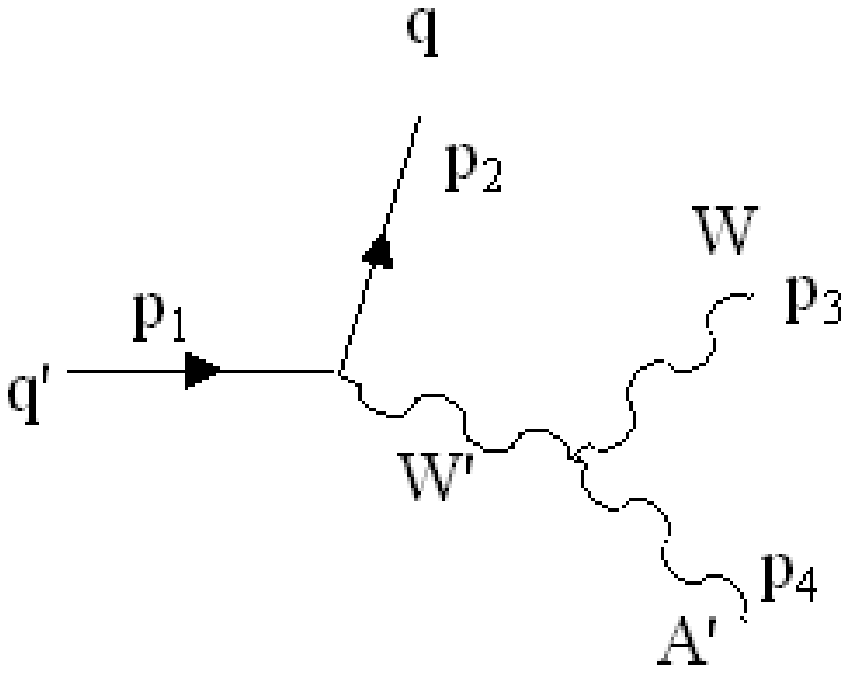}} \\
&=& g_2 \bar{u}(p_2)\gamma_\mu P_L u(p_1) \frac{i}{q^2-m^2_{W^{\prime}}}\left(-g^{\mu\nu} + \frac{q^\mu
q^\nu}{m^2_{W^{\prime}}} \right)  \\
&\times& e\left(g_{\rho\nu} (-q-p_3)_\sigma + g_{\nu\sigma}(p_4+q)_\rho + g_{\sigma\rho}(p_3-p_4)_\nu\right)
\epsilon^\sigma(p_4) \epsilon^\rho(p_3)
\end{eqnarray*}
We have borrowed the usual Standard Model coupling for the vertices. The probability amplitude is then,
\begin{equation*}
\sum_{pol} |\mathcal{M}|^2 = c_2 (t_{qW})^2 + c_1 t_{qW} + c_0
\end{equation*}
Where the coefficient functions are,
\begin{eqnarray*}
c_2 &=& \frac{1}{(q^2-m^2_{W^{\prime}})^2} \frac{m_W^4 + (m_{A^{\prime}}^2 -q^2)^2 + 2m_W^2(5 m_{A^{\prime}}^2
-q^2)}{m_W^2 m_{A^{\prime}}^2} \\
\\
c_1 &=& \frac{1}{(q^2-m^2_{W^{\prime}})^2} \frac{1}{m^2_{W^{\prime}} m^2_{A^{\prime}} m_W^2}\\
&\times& (-(m_{q^{\prime}}^2 - m_q^2)(m_W^2 - m_{A^{\prime}}^2)(m_{A^{\prime}}^4 + 10 m_W^2 m_{A^{\prime}}^2 +
m_{A^{\prime}}^4 - q^4) \\
&+ & m_{W^{\prime}}^2(-m_W^6 - 11 m_W^4 m_{A^{\prime}}^2 - 11 m_W^2 m_{A^{\prime}}^4 - m_{A^{\prime}}^6 + 3 m_W^4 q^2 \\
&+& 14 m_W^2 m_{A^{\prime}}^2 q^2 + 3 m_{A^{\prime}}^4 q^2 - 3 m_W^2 q^4 -3 m_{A^{\prime}}^2 q^4 + q^6
\\&+&
m_{q^{\prime}}^2 (m_W^4 - 10 m_W^2 m_{A^{\prime}}^2 - 3 m_{A^{\prime}}^4 + 4 m_{A^{\prime}}^2 S -q^4)) \\
&+& (m_q^2 (-3 m_W^4 + m_{A^{\prime}}^4 - q^4 + m_W^2(-10m_{A^{\prime}}^2 + 4q^2))))
\end{eqnarray*}
The last term, $c_0$ is too complicated to present and carries little significance. It is not hard to show that $c_2$
is always positive for any real choice of $q^2$.

\begin{figure}
\begin{center}
\includegraphics[scale=0.5]{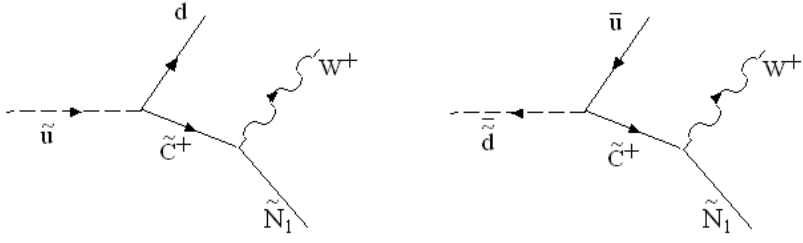}
\end{center}
\caption{Two different process contributing to squark decay into $\tilde{C}^+$ and a quark. The chargino consequently
decays into $W^\pm$ and LSP.} \label{fig: AppsqrkD}
\end{figure}

Next we are interested in the corresponding SUSY decay chain shown in Fig. (\ref{fig: AppsqrkD}). The first vertex is
given by,
\begin{equation}
  -g_2V_{11}~ \bar{d} P_R \tilde{C}^c ~\tilde{u} \quad \mathrm{and} \quad    -g_2 U^*_{11}~\bar{\tilde{C}} P_L u
  ~\tilde{d}^*
\end{equation}
respectively. The $\tilde{N}\tilde{C}W^+$ vertex is given by,
\begin{equation}
  g_2 W_\mu^- \tilde{\bar{N}}_i\gamma^\mu (\mathcal{O}^L_{ij} P_L + \mathcal{O}^R_{ij}P_R) \tilde{C}_j
\end{equation}
The matrix element for the decay initiated by a down-type squark is,
\begin{eqnarray}
  \mathcal{M}_{\tilde{\bar{d}}} &=& g_2^2 U^*_{11} ~ \bar{u}_{\tilde{N}}(p_4) \gamma^\mu (\mathcal{O}^L_{11} P_L +
  \mathcal{O}^R_{11}P_R) \frac{\slashed{q} + M_{\tilde{C}}}{q^2-M_{\tilde{C}}^2} P_L v_u(p_2) \epsilon_\mu(p_3) \\ &=& U_{11}^*~\bar{u}_{\tilde{N}}(p_4)
  \gamma^\mu\frac{ a \slashed{q} + b M_{\tilde{C}}}{q^2-M_{\tilde{C}}^2} v_u(p_2) \epsilon_\mu(p_3)
\end{eqnarray}
where,
\begin{equation}
  a = g_2^2 O_{11}^R \quad \mathrm{and} \quad b = g_2^2 O_{11}^L
\end{equation}
The squared amplitude is given by,
\begin{equation}
  |\mathcal{M}_{\tilde{\bar{d}}}|^2 = \frac{(b^2 m^2_{{\tilde{C}}} - a^2 q^2)(q^2-2m_W^2+m_{{\tilde{N}}}^2)}{2m_W^2(q^2 - m^2_{\tilde{C}})^2}~
  t_{qW} + \mathrm{Const}.
\end{equation}
Therefore, in the narrow width limit, the slope depends on the difference $b^2-a^2$. The problem is that the second
diagram contributing to this process (with an up-squark decay) has the opposite sign for this coefficient. Notice that,

\begin{eqnarray}
\bar{d} P_R {\tilde{C}}^c \tilde{u} &=& -(\tilde{C}^c)^T P_R \bar{d}^T \tilde{u}= -\bar{\tilde{C}}C^T P_R \bar{d}^T
\tilde{u}
\\\nonumber &=& -\bar{\tilde{C}} P_R C^T \bar{d}^T \tilde{u}= -\bar{\tilde{C}} P_R C^T \bar{d}^T \tilde{u} \\ \nonumber &=&
\bar{\tilde{C}} P_R  d^c \tilde{u}
\end{eqnarray}

When contracting this operator with the $\tilde{N}\tilde{C}W^+$ vertex we get the opposite spin structure,
\begin{eqnarray}
\mathcal{M}_{\tilde{u}} &=& g_2^2 V_{11} ~ \bar{u}_{\tilde{N}}(p_4) \gamma^\mu (\mathcal{O}^L_{11} P_L +
\mathcal{O}^R_{11}P_R) \frac{\slashed{q} + M_{\tilde{C}}}{q^2-M_{\tilde{C}}^2} P_R v_d(p_2) \epsilon_\mu(p_3)
\\\nonumber &=& V_{11}~ \bar{u}_{\tilde{N}}(p_4) \gamma^\mu\frac{ b \slashed{q} + a M_{\tilde{C}}}{q^2-M_{\tilde{C}}^2} v_u(p_2)
\epsilon_\mu(p_3)
\end{eqnarray}
Experimentally, we cannot distinguish between up quarks and down quarks. We must therefore average over the two
contributions,
\begin{equation}
\sum_{\tilde{u},\tilde{d}}|\mathcal{M}|^2 \propto (b^2-a^2) (f_d (U_{11}^*)^2 -f_u (V_{11})^2) t_{qW} + \mathrm{Const}.
\end{equation}
$f_d$ and $f_u$ is the fraction of events with a down-squark or up-squark, respectively.

\renewcommand{\theequation}{B-\arabic{equation}}
\setcounter{equation}{0}
\section{HERWIG implementation}
\label{app: HERWIG}

In this appendix we review the implementation of matrix elements into HERWIG. This section is relevant to any
Monte-Carlo program using the $S$ and $F$ functions of Eijk and Kleiss \cite{vanEijk:1990zp} to calculate helicity
amplitudes. The usage of these functions results in very efficient computations. The price the user has to pay is the
complexity of the expressions. These functions are then used in the spin correlations algorithm devised by Knowles and
Collins \cite{Collins:1987cp,Knowles:1988vs}. We hope to provide a brief but fairly self-consistent presentation below.

HERWIG is an event generator consisting of roughly 5 phases (for a complete description of the program consult
\cite{Corcella:2002jc}):
\begin{enumerate}
\item Hard process, where the particles in the main $2\rightarrow 2,3$ event are generated, e.g. $q~q\rightarrow g~g$ or $e^+e^-\rightarrow q \bar{q}$.
\item The parton shower phase involving the QCD evolution from the collision energy to the infrared cutoff.
\item Decay of heavy unstable particles before hadronization, such as top quark and SUSY partners.
\item Hadronization stage
\item Decay of unstable hadrons.
\end{enumerate}

We are mainly concerned with the third step. In order to decay any unstable particle (e.g. gluino, squark etc.), one
must provide HERWIG with its different decay channels and the corresponding matrix elements. To keep track of spin
correlations the matrix element must include the external polarizations. For example, let's consider the decay of a
heavy fermion into a fermion and a scalar (top decay into bottom and higgs). The matrix element is given by,

\begin{equation}
  \label{AppA eqn:ME1} i\mathcal{M} = i g \bar{u}(q,\lambda_2) u(p,\lambda_1)
\end{equation}
In order to compute this spinor product, HERWIG requires the user to express it in terms of the $S$ and $F$ functions
of Eijk and Kleiss \cite{vanEijk:1990zp} defined below. This facilitates the algebra, but obscures the expression.
Let's briefly recall the construction.

One begins by expressing any polarization in terms of massless spinors (for a pedagogical review,
see\cite{Kleiss:1985yh}). Two basic 4-momenta, $k_0, k_1$ are chosen such that,
\begin{equation}
k_0\cdot k_0 = 0, k_0\cdot k_1 = 0, k_1\cdot k_1 = -1
\end{equation}

 The basic left and right helicities are then defined via,
 \begin{equation}
   u_L(k_0)\bar{u}_L(k_0) = P_L \slashed{k}_0, \quad u_R(k_0) = \slashed{k}_1 u_L(k_0)
 \end{equation}
The helicity spinor for any other momenta (not collinear with $k_0$) is then given by,
\begin{equation}
  u_\lambda(k) = \slashed{k} u_{-\lambda}(k_0) /\sqrt{2k_0\cdot k}
\end{equation}

A massive spinor is nothing but the linear combination of two massless spinors of opposite helicities. It can be
written as,
\begin{eqnarray}
\nonumber u(p,\lambda) &=& \frac{(\slashed{p} + m)~u_{-\lambda}(k_2)}{\sqrt{2p\cdot k_2}}\\
&=& \frac{1}{\sqrt{2p\cdot k_2}}\left(S_{\lambda}(k_1,k_2)~ u_\lambda(k_1) + m~ u_{-\lambda}(k_2)\right)
\end{eqnarray}
where $p=k_1+k_2$ is decomposed into two massless 4-vectors and the $S$ function is defined as,
\begin{equation}
  S_\lambda(k_1,k_2) = \bar{u}_\lambda(k_1) u_{-\lambda}(k_2) = (S_{-\lambda}(k_2,k_1))^* = -S_\lambda(k_2,k_1)
\end{equation}
Notice that, $|S_\lambda(k_1,k_2)|^2 = m^2$.

All expressions can therefore be reduced into products of massless spinors, with the possibility of having a
$\gamma^\mu$ matrix sandwiched in between. It proves useful to define the $F$ function as well,
 \begin{equation}
F(k_1,\lambda_1,p,k_2,\lambda_2,M) = \bar{u}_{\lambda_1}(k_2)~(\slashed{p} + M)~u_{\lambda_2}(k_2)
\end{equation}
for some $p$ which is not light-like. Matrix elements for many typical processes were already implemented in HERWIG
by P. Richardson. As an example, the above matrix element (\ref{AppA eqn:ME1}) can be written as,
\begin{eqnarray}
\nonumber \mathcal{M} &=& g \bar{u}(q,\lambda_2) u(p,\lambda_1) \\ \nonumber &=& g \frac{1}{\sqrt{2p\cdot
p_2}}\frac{1}{\sqrt{2q\cdot q_2}}\left(S^*_{\lambda_2}(q_1,q_2) \bar{u}_{\lambda_2}(q_1) + m_2
\bar{u}_{-\lambda_2}(q_2)\right)\left( \slashed{p} + m_1\right)u_{-\lambda_1}(p_2) \\\nonumber \\\nonumber &=&
\frac{1}{\sqrt{2p\cdot p_2}}\frac{1}{\sqrt{2q\cdot q_2}} \\ \nonumber  &\times & g \left(S_{-\lambda_2}(q_2,q_1)
F(q_1,\lambda_2,p,p_2,-\lambda_1,m_1) + m_2 F(q_2,-\lambda_2,p,p_2,-\lambda_1,m_1) \right)
\end{eqnarray}

For a massless spin-1 particle the external polarization $\epsilon^\mu(k,\lambda)$ can be expressed as,
\begin{equation}
\label{AppA eqn:pol} \epsilon^\mu(k,\lambda) = \bar{u}_\lambda(k)\gamma^\mu u_\lambda(k_1)/\sqrt{4 k\cdot k_1}
\end{equation}
where $k_1$ is any light-like momentum not collinear with $k$.

The extra polarization of a massive spin-1 particle adds an extra complication to the calculation. Since the only
massive gauge bosons in the MSSM are the $W^\pm$ and $Z$, it is easier to simply insert the entire matrix element (e.g.
$t\rightarrow b+e^+ + \nu_e$ rather than $t\rightarrow b + W^+$ and $W^+\rightarrow e^+ + \nu_e$). In Same-Spin
theories, there are many massive gauge bosons around and it proves useful to develop the needed formalism and implement
it in HERWIG.

Looking at the massless polarization (\ref{AppA eqn:pol}) it is easy to guess the form of a massive one in terms of
spinors,
\begin{equation}
 \label{AppB eqn: pol} \epsilon^\mu(p,\lambda_1,\lambda_2) =\frac{1}{ (2\sqrt{2} m)}~ \bar{u}(p,\lambda_1)\gamma^\mu v(p,\lambda_2)
\end{equation}
where, $p^2=m^2$ and $\lambda_{1,2}$ are the usual spinor polarizations. This might seem wrong at first, as it seems to
imply 4 polarizations rather than the required 3, but as we will see in a moment, $(+,+)$ and $(-,-)$ both correspond
to the scalar polarization. First, let's verify that this indeed reproduces the correct polarization sum,
\begin{eqnarray}
\nonumber \sum_{\lambda_1,\lambda_2} \epsilon_\mu(p,\lambda_1,\lambda_2) \epsilon^*_\nu(p,\lambda_1,\lambda_2)/(8m^2)
&=& \sum_{\lambda_1,\lambda_2} Tr( u(p,\lambda_1) \bar{u}(p,\lambda_1)\gamma_\mu v(p,\lambda_2)\bar{v}(p,\lambda_2) \gamma_\nu )/(8m^2) \\
&=& Tr((\slashed{p}+m)\gamma_\mu(\slashed{p}-m)\gamma_\nu)/(8m^2) \\ &=& \left(-g_{\mu\nu} + \frac{p_\mu
p_\nu}{m^2}\right)
\end{eqnarray}
It is straight forward to show that (\ref{AppB eqn: pol}) corresponds to the different polarizations directly. First
note that,
\begin{eqnarray*}
&\epsilon_\mu&(p,\lambda_1,\lambda_2) = \frac{1}{2\sqrt{2}m}\frac{1}{2p\cdot p_1} \\&\times&
\left(S^*_{\lambda_1}(p_1,p_2)\bar{u}_{\lambda_1}(p_1) + m~\bar{u}_{-\lambda_1}(p_2)\right)\gamma_\mu
\left(S_{\lambda_2}(p_1,p_2)u_{\lambda_2}(p_1) - m ~ u_{-\lambda_2}(p_2)\right) \\ &=& \begin{cases}
\bar{u}_\lambda(p_1)\gamma_\mu u_\lambda(p_1) - \bar{u}_\lambda(p_2)\gamma_\mu u_\lambda(p_2)~ /2\sqrt{2}m & \lambda_1
\equiv \lambda =\lambda_2 \\ \frac{\left(S_{-\lambda}-S^*_\lambda \right)}{m}~ \bar{u}_\lambda(p_2)\gamma_\mu
u_\lambda(p_1)~/2\sqrt{2}m & \lambda_1 \equiv \lambda \ne \lambda_2 \end{cases}
\end{eqnarray*}

In the rest frame of the particle we can take $\vec{p}_1 = \hat{z}|p|/2$ and  $\vec{p}_2 = -\hat{z}|p|/2$, so that $p =
p_1+p_2 = (m,\vec{0})$ and the massless spinors are given by,
\begin{equation}
u_-(p_1) = \sqrt{2m}\left(%
\begin{array}{c}
  0 \\
  1 \\
  0 \\
  0 \\
\end{array}%
\right)
\quad
u_+(p_1) = \sqrt{2m}\left(%
\begin{array}{c}
  0 \\
  0 \\
  1 \\
  0 \\
\end{array}%
\right)
\end{equation}
With a few lines of arithmetic one can verify that,
\begin{eqnarray*}
\epsilon_\mu(p,-,-) &=& \left((1,0,0,m) - (1,0,0,-m)\right)/(2\sqrt{2}m) = (0,0,0,1/\sqrt{2}) \\
\epsilon_\mu(p,+,+) &=& \left((1,0,0,m) - (1,0,0,-m)\right)/(2\sqrt{2}m) = (0,0,0,1/\sqrt{2}) \\
\epsilon_\mu(p,-,+) &=& 2(0,m,i m,0)/(2\sqrt{2}m) = (0,1/\sqrt{2},i/\sqrt{2},0) \\
\epsilon_\mu(p,+,-) &=& 2(0,m,-i m,0)/(2\sqrt{2}m) = (0,1/\sqrt{2},-i/\sqrt{2},0)
\end{eqnarray*}
and so we conclude that $\epsilon^\mu(-,+) =\epsilon^\mu_L$, $\epsilon^\mu(+,-) =\epsilon^\mu_R$ and $\epsilon^\mu(-,-)
= \epsilon^\mu(+,+) = \epsilon^\mu_0/\sqrt{2}$. It is now straight forward, albeit tedious, to express any matrix
element involving an external massive gauge boson in terms of $S$ and $F$ functions and implement it into HERWIG.

We begin with the matrix element for a gauge boson decay into a fermion - anti-fermion pair,
\begin{equation}
i\mathcal{M} = i A_\lambda \bar{u}(k,\lambda_2) \gamma^\mu P_\lambda v(q,\lambda_4) ~~\epsilon_\mu(p,\lambda_1,\lambda_2)
\end{equation}
The corresponding expressions are,
\begin{eqnarray*}
\mathcal{M}((\lambda_1,\lambda_1),\lambda_3,\lambda_4) &=& \frac{\sqrt{2}}{\sqrt{4 p\cdot p_2 2 k\cdot k_2 2 q \cdot q_2}} \\
&\times &\left( A_{-\lambda_1}\left(F(k_2,-\lambda_3,k,p_1,\lambda_1,m_2)~F(p_1,\lambda_1,q,q_2,-\lambda_4,-m_3) +
(p_1\rightarrow p_2)\right)\right. \\
&+& \left. (\lambda_1 \rightarrow -\lambda_1) \right)\\
\end{eqnarray*}
\begin{eqnarray*}
\mathcal{M}((\lambda_1,-\lambda_1),\lambda_3,\lambda_4) &=& \frac{1}{\sqrt{4 p\cdot p_2 2 k\cdot k_2 2 q \cdot q_2}} \\
&\times & \left(\frac{S_{-\lambda_1}(p_1,p_2) - S^*_{\lambda_1}(p_1,p_2)}{m_1}\right)\\ &\times &
\left(~A_{-\lambda_1} F(k_2,-\lambda_3,k,p_2,\lambda_1,m_2)~F(p_1,\lambda_1,q,q_2,-\lambda_4,-m_3)\right. \\&~&\left.
~~+ (p_1\leftrightarrow p_2, \lambda_1\rightarrow -\lambda_1)~~\right)
\end{eqnarray*}
The factor of $\sqrt{2}$ above is to guarantee proper normalization of the longitudinal mode. With a little bit of care
it is easy to obtain the other two diagrams $f(\bar{f})\rightarrow f(\bar{f}) + g.b.$. To turn a fermion into an
anti-fermion or vice-versa simply send $m\rightarrow -m$. The incoming gauge-boson polarization becomes an outgoing one
by conjugation,
\begin{equation}
\label{AppB eqn: inc out} \epsilon_\mu \rightarrow \epsilon_\mu^* = \bar{v}(p,\lambda_2)\gamma_\mu u(p,\lambda_1) =
\lambda_1\lambda_2 ~ \bar{u}(p,-\lambda_1) \gamma_\mu v(p,-\lambda_2)
\end{equation}
So we simply have to send $\lambda_1\rightarrow -\lambda_1$ in the expressions above, and an overall minus sign in
front of the $\mathcal{M}((\lambda_1,-\lambda_1),\lambda_3,\lambda_4)$ amplitude due to the $\lambda_1\lambda_2$
factor.

The last diagram we consider is the non-Abelian vertex including 3 external massive spin-1 polarizations.
\begin{eqnarray}
\nonumber i\mathcal{M} &=& \vcenter{
\includegraphics[scale=0.5]{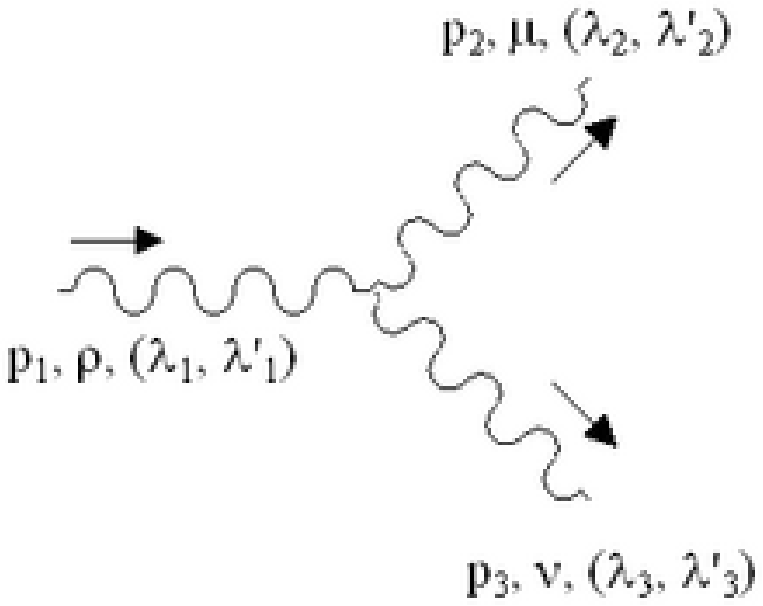}} \\
&=& i g \left( g^{\mu\nu}(p_3-p_2)^\rho + g^{\nu\rho}(-p_1-p_3)^\mu + g^{\rho\mu}(p_2+p_1)^\nu \right) \\
&\times& \epsilon_\rho(p_1,\lambda_1,\lambda'_1) ~ \epsilon^*_\mu (p_2,\lambda_2,\lambda'_2) ~
\epsilon^*_\nu(p_3,\lambda_3,\lambda'_3)
\end{eqnarray}
We give the corresponding expression for the first term only. The other terms can be obtained by trivial permutations
and using (\ref{AppB eqn: inc out}) for the outgoing polarizations.
\begin{eqnarray}
i\mathcal{M}& \supset& \epsilon^*_\mu (p_2,\lambda_2,\lambda'_2) \epsilon^*_\nu(p_3,\lambda_3,\lambda'_3)
g^{\mu\nu}(p_3-p_2)^\rho \epsilon_\rho(p_1,\lambda_1,\lambda'_1) \\\nonumber &=&\frac{1}{16\sqrt{2} m_1m_2m_3}
A(\lambda_2,\lambda'_2,\lambda_3,\lambda'_3) B(\lambda_1,\lambda'_1)
\end{eqnarray}
where,
\begin{equation*}
A(\lambda_2,\lambda'_2,\lambda_3,\lambda'_3) =
\begin{cases}
\lambda_2 = \lambda'_2 &\begin{array}{c}
                          \frac{1}{m_3^2}\left(F(q'_3,\lambda_3,p_3,q_2,\lambda_2,m_3) F(q_2,\lambda_2,p_3,q'_3,\lambda'_3,-m_3)
                          \right.\\
                          - F(q'_3,\lambda_3,p_3,q'_2,\lambda_2,m_3) F(q'_2,\lambda_2,p_3,q'_3,\lambda'_3,-m_3) \\
                          +\left. \lambda_3 \rightarrow -\lambda_3\right)
                        \end{array} \\
 & \\
\lambda_2 = -\lambda'_2 & \begin{array}{c}
                           \frac{1}{m_3^2}\left(\frac{S(q_2,q'_2,-\lambda_2) - S^*(q_2,q'_2,\lambda_2)}{m_2} \right) \times \\
                           \left(F(q'_3,\lambda_3,p_3,q'_2,\lambda_2,m_3)F(q_2,\lambda_2,p_3,q'_3,\lambda'_3,-m_3) \right.\\
                           \left. +(\lambda_3\rightarrow -\lambda_3, q'_2 \leftrightarrow q_2) \right)
                          \end{array}
\end{cases}
\end{equation*}
and,
\begin{equation*}
B(\lambda_1,\lambda'_1) =
\begin{cases}
\lambda_1 = \lambda'_1 & \left(F(q_1,\lambda_1,p_3-p_2,q_1,\lambda_1,0) -
F(q'_1,-\lambda_1,p_3-p_2,q'_1,-\lambda_1,0)\right) \\
& \\
\lambda_1 = -\lambda'_1 & \left(\frac{S(q_1,q'_1,-\lambda_1)-S^*(q_1,q'_1,\lambda_1)}{m_1} F(q_1,\lambda_1,p_3-p_2,
q'_1,\lambda_1,0) \right)
\end{cases}
\end{equation*}
Here, the momenta is written in terms of two massless momenta $p_i = q_i + q'_i$, with $q_i^2 = (q'_i)^2 = 0$ and
$p_i^2 = m_i^2$.

\bibliography{SpCrLHC}

\providecommand{\href}[2]{#2}\begingroup\raggedright\begin{thebibliography}{10}

\bibitem{Baer:1995nq}
H.~Baer, C.-h. Chen, F.~Paige, and X.~Tata, {\it Signals for minimal
  supergravity at the cern large hadron collider: Multi - jet plus missing
  energy channel},  {\em Phys. Rev.} {\bf D52} (1995) 2746--2759,
  [\href{http://xxx.lanl.gov/abs/hep-ph/9503271}{{\tt hep-ph/9503271}}].

\bibitem{Baer:1995va}
H.~Baer, C.-h. Chen, F.~Paige, and X.~Tata, {\it Signals for minimal
  supergravity at the cern large hadron collider ii: Multilepton channels},
  {\em Phys. Rev.} {\bf D53} (1996) 6241--6264,
  [\href{http://xxx.lanl.gov/abs/hep-ph/9512383}{{\tt hep-ph/9512383}}].

\bibitem{Hinchliffe:1996iu}
I.~Hinchliffe, F.~E. Paige, M.~D. Shapiro, J.~Soderqvist, and W.~Yao, {\it
  Precision susy measurements at lhc},  {\em Phys. Rev.} {\bf D55} (1997)
  5520--5540, [\href{http://xxx.lanl.gov/abs/hep-ph/9610544}{{\tt
  hep-ph/9610544}}].

\bibitem{Hinchliffe:1998zj}
I.~Hinchliffe {\em et~al.}, {\it Precision susy measurements at lhc: Point 3},
  . LBNL-40954.

\bibitem{unknown:1999fr}
{\it Atlas detector and physics performance. technical design report. vol. 1
  and 2}, . CERN-LHCC-99-15.

\bibitem{Lester:1999tx}
C.~G. Lester and D.~J. Summers, {\it Measuring masses of semi-invisibly
  decaying particles pair produced at hadron colliders},  {\em Phys. Lett.}
  {\bf B463} (1999) 99--103,
  [\href{http://xxx.lanl.gov/abs/hep-ph/9906349}{{\tt hep-ph/9906349}}].

\bibitem{Lester:2005je}
C.~G. Lester, M.~A. Parker, and M.~J. White, {\it Determining susy model
  parameters and masses at the lhc using cross-sections, kinematic edges and
  other observables},  {\em JHEP} {\bf 01} (2006) 080,
  [\href{http://xxx.lanl.gov/abs/hep-ph/0508143}{{\tt hep-ph/0508143}}].

\bibitem{Miller:2005zp}
D.~J. Miller, P.~Osland, and A.~R. Raklev, {\it Invariant mass distributions in
  cascade decays},  {\em JHEP} {\bf 03} (2006) 034,
  [\href{http://xxx.lanl.gov/abs/hep-ph/0510356}{{\tt hep-ph/0510356}}].

\bibitem{Arkani-Hamed:2005px}
N.~Arkani-Hamed, G.~L. Kane, J.~Thaler, and L.-T. Wang, {\it Supersymmetry and
  the lhc inverse problem},  \href{http://xxx.lanl.gov/abs/hep-ph/0512190}{{\tt
  hep-ph/0512190}}.

\bibitem{Arkani-Hamed:2001ca}
N.~Arkani-Hamed, A.~G. Cohen, and H.~Georgi, {\it (de)constructing dimensions},
   {\em Phys. Rev. Lett.} {\bf 86} (2001) 4757--4761,
  [\href{http://xxx.lanl.gov/abs/hep-th/0104005}{{\tt hep-th/0104005}}].

\bibitem{Appelquist:2000nn}
T.~Appelquist, H.-C. Cheng, and B.~A. Dobrescu, {\it Bounds on universal extra
  dimensions},  {\em Phys. Rev.} {\bf D64} (2001) 035002,
  [\href{http://xxx.lanl.gov/abs/hep-ph/0012100}{{\tt hep-ph/0012100}}].

\bibitem{Cheng:2003ju}
H.-C. Cheng and I.~Low, {\it Tev symmetry and the little hierarchy problem},
  {\em JHEP} {\bf 09} (2003) 051,
  [\href{http://xxx.lanl.gov/abs/hep-ph/0308199}{{\tt hep-ph/0308199}}].

\bibitem{Cheng:2004yc}
H.-C. Cheng and I.~Low, {\it Little hierarchy, little higgses, and a little
  symmetry},  {\em JHEP} {\bf 08} (2004) 061,
  [\href{http://xxx.lanl.gov/abs/hep-ph/0405243}{{\tt hep-ph/0405243}}].

\bibitem{Low:2004xc}
I.~Low, {\it T parity and the littlest higgs},  {\em JHEP} {\bf 10} (2004) 067,
  [\href{http://xxx.lanl.gov/abs/hep-ph/0409025}{{\tt hep-ph/0409025}}].

\bibitem{Cheng:2005as}
H.-C. Cheng, I.~Low, and L.-T. Wang, {\it Top partners in little higgs theories
  with t-parity},  \href{http://xxx.lanl.gov/abs/hep-ph/0510225}{{\tt
  hep-ph/0510225}}.

\bibitem{Datta:2005vx}
A.~Datta, G.~L. Kane, and M.~Toharia, {\it Is it susy?},
  \href{http://xxx.lanl.gov/abs/hep-ph/0510204}{{\tt hep-ph/0510204}}.

\bibitem{Meade:2006dw}
P.~Meade and M.~Reece, {\it Top partners at the lhc: Spin and mass
  measurement},  \href{http://xxx.lanl.gov/abs/hep-ph/0601124}{{\tt
  hep-ph/0601124}}.

\bibitem{Barr:2004ze}
A.~J. Barr, {\it Using lepton charge asymmetry to investigate the spin of
  supersymmetric particles at the lhc},  {\em Phys. Lett.} {\bf B596} (2004)
  205--212, [\href{http://xxx.lanl.gov/abs/hep-ph/0405052}{{\tt
  hep-ph/0405052}}].

\bibitem{Datta:2005zs}
A.~Datta, K.~Kong, and K.~T. Matchev, {\it Discrimination of supersymmetry and
  universal extra dimensions at hadron colliders},  {\em Phys. Rev.} {\bf D72}
  (2005) 096006, [\href{http://xxx.lanl.gov/abs/hep-ph/0509246}{{\tt
  hep-ph/0509246}}].

\bibitem{Smillie:2005ar}
J.~M. Smillie and B.~R. Webber, {\it Distinguishing spins in supersymmetric and
  universal extra dimension models at the large hadron collider},  {\em JHEP}
  {\bf 10} (2005) 069, [\href{http://xxx.lanl.gov/abs/hep-ph/0507170}{{\tt
  hep-ph/0507170}}].

\bibitem{Alves:2006df}
A.~Alves, O.~Eboli, and T.~Plehn, {\it It's a gluino},
  \href{http://xxx.lanl.gov/abs/hep-ph/0605067}{{\tt hep-ph/0605067}}.

\bibitem{Cheng:2002ab}
H.-C. Cheng, K.~T. Matchev, and M.~Schmaltz, {\it Bosonic supersymmetry?
  getting fooled at the lhc},  {\em Phys. Rev.} {\bf D66} (2002) 056006,
  [\href{http://xxx.lanl.gov/abs/hep-ph/0205314}{{\tt hep-ph/0205314}}].

\bibitem{Allanach:2002nj}
B.~C. Allanach {\em et~al.}, {\it The snowmass points and slopes: Benchmarks
  for susy searches},  {\em Eur. Phys. J.} {\bf C25} (2002) 113--123,
  [\href{http://xxx.lanl.gov/abs/hep-ph/0202233}{{\tt hep-ph/0202233}}].

\bibitem{Corcella:2002jc}
G.~Corcella {\em et~al.}, {\it Herwig 6.5 release note},
  \href{http://xxx.lanl.gov/abs/hep-ph/0210213}{{\tt hep-ph/0210213}}.

\bibitem{Collins:1987cp}
J.~C. Collins, {\it Spin correlations in monte carlo event generators},  {\em
  Nucl. Phys.} {\bf B304} (1988) 794.

\bibitem{Knowles:1987cu}
I.~G. Knowles, {\it Angular correlations in qcd},  {\em Nucl. Phys.} {\bf B304}
  (1988) 767.

\bibitem{Knowles:1988hu}
I.~G. Knowles, {\it A linear algorithm for calculating spin correlations in
  hadronic collisions},  {\em Comput. Phys. Commun.} {\bf 58} (1990) 271--284.

\bibitem{Knowles:1988vs}
I.~G. Knowles, {\it Spin correlations in parton - parton scattering},  {\em
  Nucl. Phys.} {\bf B310} (1988) 571.

\bibitem{Richardson:2001df}
P.~Richardson, {\it Spin correlations in monte carlo simulations},  {\em JHEP}
  {\bf 11} (2001) 029, [\href{http://xxx.lanl.gov/abs/hep-ph/0110108}{{\tt
  hep-ph/0110108}}].

\bibitem{Barr:2005dz}
A.~J. Barr, {\it Measuring slepton spin at the lhc},
  \href{http://xxx.lanl.gov/abs/hep-ph/0511115}{{\tt hep-ph/0511115}}.

\bibitem{Chung:2003fi}
D.~J.~H. Chung {\em et~al.}, {\it The soft supersymmetry-breaking lagrangian:
  Theory and applications},  {\em Phys. Rept.} {\bf 407} (2005) 1--203,
  [\href{http://xxx.lanl.gov/abs/hep-ph/0312378}{{\tt hep-ph/0312378}}].

\bibitem{Barger:1987nn}
V.~D. Barger and R.~J.~N. Phillips, {\it Collider physics}, . REDWOOD CITY,
  USA: ADDISON-WESLEY (1987) 592 P. (FRONTIERS IN PHYSICS, 71).

\bibitem{Macesanu:2002hg}
C.~Macesanu, C.~D. McMullen, and S.~Nandi, {\it Collider implications of models
  with extra dimensions},  \href{http://xxx.lanl.gov/abs/hep-ph/0211419}{{\tt
  hep-ph/0211419}}.

\bibitem{Han:private}
T.~Han, {\em private communication.}

\bibitem{vanEijk:1990zp}
B.~van Eijk and R.~Kleiss, {\it On the calculation of the exact g g $\to$ z b
  anti-b cross- section including z decay and b quark mass effects}, . In
  *Aachen 1990, Proceedings, Large Hadron Collider, vol. 2* 183-194. CERN
  Geneva - CERN-90-10-B (90/12,rec.Jul.91) 183- 194.

\bibitem{Kleiss:1985yh}
R.~Kleiss and W.~J. Stirling, {\it Spinor techniques for calculating p anti-p
  $\to$ w+- / z0 + jets},  {\em Nucl. Phys.} {\bf B262} (1985) 235--262.

\end{thebibliography}\endgroup
\bibliographystyle{jhep}
\end{document}